\documentclass{IEEEtran}

\usepackage[svgnames]{xcolor}
\usepackage{graphicx}
\usepackage[font=small, skip=5pt, belowskip=0pt]{caption}
\usepackage[font=footnotesize]{subcaption}
\usepackage{multirow}
\usepackage{array}
\usepackage{booktabs}
\usepackage{pdflscape}
\usepackage[ruled,vlined,linesnumbered]{algorithm2e}
\usepackage[T1]{fontenc}
\usepackage[utf8]{inputenc}
\usepackage{cite}
\usepackage{url}
\usepackage{comment}
\usepackage{mleftright,xparse}
\usepackage{amssymb,amsmath,amsfonts,amsthm,bm,mathtools,cuted,bbold}
\usepackage{siunitx}
\newtheorem{remark}{Remark}
\newtheorem{assumption}{Assumption} 

\usepackage{tikz}
\usetikzlibrary{math, shapes, calc}
\usepackage{pgfplots}
\usepgfplotslibrary{groupplots}
\pgfplotsset{
compat=1.13,
legend style={font=\footnotesize, fill opacity=0.7,  draw opacity=1, text opacity=1, draw=white!15!black, legend cell align=left, align=left}, 
width=6cm, 
height=6cm,
yminorticks=false,
xminorticks=false,
label style={font=\small},
title style={font=\small},
tick style={color=black},
tick align=outside,
tick pos=left,
tick label style={font=\footnotesize},
grid style={line width=.1pt, draw=gray!20},
major grid style={line width=.1pt,draw=gray!20},}

\usepackage[normalem]{ulem}

\usepackage[acronym]{glossaries}
\newacronym{3d}{3D}{three dimensional}
\newacronym{aoa}{AoA}{angle of arrival}
\newacronym{aod}{AoD}{angle of departure}
\newacronym{ap}{AP}{access point}
\newacronym{awgn}{AWGN}{additive white gaussian noise}
\newacronym{b5g}{B5G}{Beyond-5G}
\newacronym{BSW}{BSW}{beam sweeping}
\newacronym{bsw}{CB-BSW}{codebook-based beam sweeping}
\newacronym[plural=BSs, firstplural=base stations (BSs)]{bs}{BS}{base station}
\newacronym{ctrl}{ctrl}{control}
\newacronym{cc}{CC}{control channel}
\newacronym{cdf}{CDF}{cumulative density function}
\newacronym{ce}{CE}{channel estimation}
\newacronym{oce}{OPT-CE}{optimization based on channel estimation}
\newacronym{obcc}{OB-CC}{out-of--band \gls{cc}} 
\newacronym{csi}{CSI}{channel state information}
\newacronym{dc}{DC}{data channel}
\newacronym{dl}{DL}{downlink}
\newacronym{dft}{DFT}{discrete Fourier transform}
\newacronym{doa}{DoA}{direction-of-arrival}
\newacronym{emf}{EMF}{electromagnetic field}
\newacronym{em}{EM}{electromagnetic}
\newacronym{fp}{FP}{fractional program}
\newacronym[plural=HRISs, firstplural=Hybrid Reconfigurable Intelligent Surfaces (HRISs)]{hris}{HRIS}{hybrid reconfigurable intelligent surface}
\newacronym{ibcc}{IB-CC}{in-band \gls{cc}}
\newacronym{ibno}{IB-no}{IB-no}
\newacronym{ibwf}{IB-wf}{IB-wf}
\newacronym{ios}{IoS}{Internet-of-Surfaces}
\newacronym{iot}{IoT}{Internet-of-Things}
\newacronym{iid}{i.i.d.}{independently identically distributed}
\newacronym[plural=KPIs, firstplural=key performance indicators (KPIs)]{kpi}{KPI}{key performance indicator}
\newacronym{lf}{LF}{low frequency}
\newacronym{los}{LoS}{line-of-sight}
\newacronym{mac}{MAC}{medium access control}
\newacronym{mcs}{MCS}{modulation and coding scheme}
\newacronym{mimo}{MIMO}{multiple-input multiple-output}
\newacronym{miso}{MISO}{multiple-input single-output}
\newacronym{ml}{ML}{machine learning}
\newacronym{mmse}{MMSE}{minimum mean squared error}
\newacronym{mrt}{MRT}{maximum-ratio transmission}
\newacronym{mse}{MSE}{mean squared error}
\newacronym{nlos}{NLoS}{non-line-of-sight}
\newacronym{ofdm}{OFDM}{orthogonal frequency-division multiplexing}
\newacronym{pdf}{pdf}{probability distribution function}
\newacronym{pla}{PLA}{planar linear array}
\newacronym{pap}{P\&P}{plug-and-play}
\newacronym{phy}{PHY}{physical-layer}
\newacronym{ppp}{PPP}{Poisson point process}
\newacronym[plural=RISs, firstplural=Reconfigurable intelligent surfaces (RISs)]{ris}{RIS}{reconfigurable intelligent surface}
\newacronym{risc}{RISC}{\gls{ris} controller}
\newacronym{rf}{RF}{radio frequency}
\newacronym{rmse}{RMSE}{root-mean-square error}
\newacronym{rss}{RSS}{received signal strength}
\newacronym{se}{SE}{spectral efficiency}
\newacronym{sdp}{SDP}{semidefinite programming}
\newacronym{sdr}{SDR}{semidefinite relaxation}
\newacronym{sinr}{SINR}{signal-to-interference-plus-noise ratio}
\newacronym{smse}{SMSE}{sum mean squared error}
\newacronym{snr}{SNR}{signal-to-noise ratio}
\newacronym{soa}{SoA}{state-of-the-art}
\newacronym{sre}{SRE}{smart radio environment}
\newacronym{toa}{ToA}{time-of-arrival}
\newacronym{tdd}{TDD}{time division multiplex}
\newacronym{tti}{TTI}{transmission time interval}
\newacronym[plural=UEs, firstplural=users equipment (UEs)]{ue}{UE}{user equipment}
\newacronym{ul}{UL}{uplink}
\newacronym{ula}{ULA}{uniform linear array}
\newacronym{urllc}{URLLC}{ultra-reliable low-latency communications}

\newcommand{\T}{^{\mathsf{T}}}     
\newcommand{\mc}[1]{\mathcal{#1}}   
\newcommand{\mb}[1]{\mathbf{#1}}    
\DeclareMathOperator*{\argmax}{arg\,max}    
\newcommand{\oce}{_\mathrm{CE}}
\newcommand{\bsw}{_\mathrm{CB}} 
\newcommand{\alg}{_\mathrm{alg}} 
\newcommand{\ini}{_\mathrm{ini}} 
\newcommand{\set}{_\mathrm{set}} 
\newcommand{\pay}{_\mathrm{pay}} 

\definecolor{amaranth}{rgb}{0.9, 0.17, 0.31}
\definecolor{cadmiumgreen}{rgb}{0.0, 0.42, 0.24}

\newcommand{\change}[1]{\textcolor{black}{#1}}

\newcommand{\fs}[1]{\textcolor{amaranth}{#1}} 

\title{\change{On the Impact of Control Signaling\\ in RIS-Empowered Wireless Communications}}

\author{
    Fabio Saggese,~\IEEEmembership{Member,~IEEE}, Victor Croisfelt,~\IEEEmembership{Student~Member,~IEEE},  Rados\l{}aw Kotaba,\\ Kyriakos Stylianopoulos,~\IEEEmembership{Student~Member,~IEEE}, George C. Alexandropoulos,~\IEEEmembership{Senior~Member,~IEEE},\\ and Petar Popovski,~\IEEEmembership{Fellow,~IEEE}\vspace{-1cm}
    \thanks{This work was partly supported by the Villum Investigator grant ``WATER'' from the Villum Foundation, Denmark, and by the EU H2020 RISE-6G project under grant number 101017011. F. Saggese, V. Croisfelt, R. Kotaba, and P. Popovski are with the Connectivity Section of the Department of Electronic Systems, Aalborg University, Aalborg, Denmark (e-mails: \{fasa, vcr, rak, petarp\}@es.aau.dk). K. Stylianopoulos and G. C. Alexandropoulos are with the Department of Informatics and Telecommunications, National and Kapodistrian University of Athens, Panepistimiopolis Ilissia, 15784 Athens, Greece (e-mails: \{kstylianop, alexandg\}@di.uoa.gr).}
}

\begin{document}

\maketitle

\begin{abstract}
    The research on Reconfigurable Intelligent Surfaces (RISs) has dominantly been focused on physical-layer aspects and analyses of the achievable adaptation of the wireless propagation environment. Compared to that, questions related to system-level integration of RISs have received less attention. 
    \change{We address this research gap by analyzing the necessary control/signaling operations that are necessary to integrate RIS as a new type of wireless infrastructure element.} We build a general model for \change{evaluating the impact of control operations \change{along two dimensions}: \emph{i}) the allocated bandwidth of the control channels (in-band and out-of-band), and \emph{ii}) the rate selection for the data channel (multiplexing or diversity).} Specifically, the second dimension results in two \change{generic} transmission schemes, one based on channel estimation and the subsequent optimization of the RIS, while the other is based on sweeping through predefined RIS phase configurations. {We analyze the communication performance in multiple setups built along these two dimensions.}
    While necessarily simplified, our analysis reveals the basic trade-offs in RIS-assisted communication and the associated control operations. \change{The main contribution of the paper is a methodology for systematic evaluation of the control overhead in RIS-aided networks, regardless of the specific control schemes used.}
\end{abstract}

\section{Introduction} \label{sec:intro}
\Glspl{ris} constitute a promising technology for sixth generation (6G) wireless networks, which has received significant attention within the relevant research community in recent years~\cite{Huang2019}. The main underlying idea is to electronically tune the reflective properties of an \gls{ris} to manipulate the phase, amplitude, and polarization of the incident electromagnetic waves~\cite{risTUTORIAL2020}. This creates a propagation environment that is, at least, partially controllable~\cite{RISE6G_COMMAG}. \Glspl{ris} can be fabricated with classical antenna elements controlled through switching elements or, more advanced, with metamaterials having tunable electromagnetic properties~\cite{EURASIP_RIS}. In the 6G context, the \gls{ris} technology has been identified as one of the cost-effective solutions to address the increasing demand for higher data rates, reduced latency, and increased coverage. In particular, an \gls{ris} can improve the received signal strength and minimize interference by reflecting signals to intended receivers and away from non-intended ones; this leads to applications aiming for increased communication security~\cite{RIS_security_mal} and/or reduced electromagnetic field exposure~\cite{NA2021}. \glspl{ris} can also extend the coverage of wireless communication systems by reflecting the signals to areas that are difficult to reach using conventional means~\cite{Croisfelt2023}.

The dominant part of the literature concerning \gls{ris}-aided communication systems deals with \gls{phy} challenges. Recent studies have explored physics-based derivation of \gls{ris}-parametrized end-to-end channel models, incorporating causality, frequency selectivity, as well as any arising mutual coupling effects~\cite{PhysFad}. 
Many other papers have investigated the potential benefits of \gls{ris}-aided systems in terms of spectral and energy efficiencies by optimizing the parameters of the \gls{ris}' elements alone or jointly with the operations of the \gls{bs} (see, \emph{e.g.},~\cite{Tsinghua_RIS_Tutorial, massive_RIS}). \change{To enable such strategies, several works have focused on designing and evaluating \emph{\gls{ce} methods in the presence of \glspl{ris}, either relying on the observed equivalent end-to-end channel from the \gls{bs} to the \gls{ue}}, when dealing with solely reflective \glspl{ris}~\cite{CascadeCE_ProcIEEE, Mo2023ce}, or directly estimating individual channels -- \gls{bs}-to-\gls{ris} and \gls{ris}-to-\gls{ue} -- by using simultaneous reflecting and sensing \glspl{ris}~\cite{CE_HRIS_2023}. The latter \gls{ris} design belongs to the attempts to minimize the \gls{ce} overhead~\cite{HRIS_VTM}, which can be considerably large due to the expected high numbers of \gls{ris} elements~\cite{popov2021experimental} or hardware-induced non-linearities~\cite{Tsinghua_RIS_Tutorial}.} A different research direction bypasses explicit \gls{ce} and relies on \emph{\gls{bsw} methods}~\cite{singh2021fast}. Accordingly, the \gls{ris} is scheduled to progressively change among reflecting configurations from a predefined codebook so that the end-to-end system can discover the most suitable configuration~\cite{RISsweeping_2020, Jamali2022, alexandropoulos2022hierarchical, An2022}. \change{The predefined codebook can be practically optimized for different purposes~\cite{rahal_RISbeams}: a suitable approach is to use hierarchical codebook structures~\cite{HierarhicalCodebook, alexandropoulos2022hierarchical}.}

Within the existing research literature, the questions related to link/\gls{mac} protocol and system-level integration of \glspl{ris} have received much less attention than \gls{phy} topics. \change{\emph{Specifically, the literature lacks a systematic treatment of the control signaling aspects, required to remotely manage the behavior of the \glspl{ris} and the \glspl{ue}. Control signaling is often exerted by the \gls{bs} and relies upon well-defined \glspl{cc}.} The study of those procedures is central to integrating \glspl{ris} as a new type of network element within the existing wireless infrastructure. In this regard, we need to understand how the system performance is impacted by the design of the \glspl{cc} in terms of its rate, reliability as well as the overhead/trade-offs incurred by the control signaling procedures. To the best of our knowledge, this paper marks the initial attempt to systematically investigate the influence of control signaling on the performance of \gls{ris}-aided wireless systems.}

\subsection{Related literature}
\change{Most of the existing literature assumes that the control over an \gls{ris}-aided communication system is error-free and instantaneous. For example, in~\cite{Croisfelt2022randomaccess} and later~\cite{Croisfelt2023}, the authors presented a detailed protocol for random access aided by an \gls{ris}. It was showcased that, despite the physical overhead of switching its configurations, the \gls{ris} brings notable performance benefits, allowing more \glspl{ue} to access the \gls{bs} on average. However, these works ignored the impact of control signaling.} Similarly, in~\cite{PHY_Retransmission}, the effect of re-transmission protocols in \gls{ris}-aided systems in the case of erroneous transmission was studied, but the control impact was ignored.

\change{Other works did not even specify the required control information to be exchanged between the communication entities. For instance, one of the first works focusing on fast \gls{ris} programmability~\cite{RISsweeping_2020} presented a multi-stage \gls{bsw} protocol. By tasking the \gls{ris} to dynamically illuminate the area where a \gls{ue} is located, the authors of~\cite{Jamali2022} introduced a \gls{dl} transmission protocol, including \gls{ue} localization, \gls{ris} configuration, and pilot-aided end-to-end \gls{ce}.} In~\cite{alexandropoulos2022hierarchical}, a fast near-field alignment scheme was proposed for the \gls{ris} phase-shifts and the transceiver beamformer, relying on a variable-width hierarchical \gls{ris} phase configuration codebook. Recently,~\cite{An2022} discussed the overhead and challenges of integrating the \gls{ris} into the network,  arguing that the reduced overhead offered by \gls{bsw} schemes benefits the overall system performance. Nevertheless, the control signaling that needs to be exchanged for those schemes was not investigated.

\subsection{Contributions}\label{sec:intro:contributions}
\change{This paper aims to introduce a model quantifying the impact of control signaling on the performance of \gls{ris}-enabled wireless communications.}
The number of actual control options is subject to a combinatorial explosion due to the system's large number of configurable parameters, such as frame size or feedback design. Obviously, we cannot address all these designs in a single work, but what we are striving for is to get a simple\change{, yet generic, model for analyzing the impact of control that captures the essential design trade-offs and can be expanded to analyze other, more elaborate designs.} \change{For this reason, in this paper, we keep the complexity of the \gls{ris}-aided system model at the minimum, focusing on analyzing and evaluating the interaction between the control and the data planes. Considering a single-antenna \gls{bs} and narrowband communications, we analyze the communication performance when control signaling occurs over the \gls{ris}-\gls{cc} linking the \gls{bs} to the \gls{ris}, and the \gls{ue}-\gls{cc} that links the \gls{bs} to the \gls{ue} through the \gls{ris}. To generalize the proposed model, we further discuss, at the end of the paper, how to relax some of the assumptions made.} We build our generic control model along the two following dimensions that capture relevant trade-offs to be studied theoretically and which can be met in practice when deploying an \gls{ris}-aided communication system.

The \emph{first dimension} is shaped around evaluating the distinctions between traditional communication paradigms of multiplexing and diversity~\cite{Popovski2020}. In a \emph{multiplexing-oriented transmission}, \change{a \gls{kpi} of interest -- \emph{e.g.}, data rate --} is adapted to the actual channel conditions based on the \gls{csi} obtained via \gls{ce}. However, \gls{ce} \change{generally needs complex control signaling, implying a high control overhead.} \change{In an \gls{ris}-aided system, for example, a multiplexing transmission corresponds to a case in which the \gls{ris} configuration is purposefully configured to maximize the \gls{snr} of the cascaded end-to-end channel, based on acquired \gls{csi}, and the data rate is chosen accordingly. In a \emph{diversity transmission} scenario, the \gls{kpi} is pre-established, and the \gls{ue} relies on the expectation that the propagation environment will accommodate it. However, if the propagation environment does not meet these expectations, it can lead to an outage event, resulting in transmission failure.} \change{In an \gls{ris}-aided system, diversity transmission can be realized by using a \gls{bsw} procedure: while the \gls{ue} transmits, the \gls{bs} commands the \gls{ris} to successively load configurations, hoping that one of them will likely support the predefined \gls{kpi}.} \change{We refer to the paradigm of \gls{ris}-aided multiplexing transmission as \emph{\gls{oce}}, while \emph{\gls{bsw}} refers to the diversity paradigm.}

\change{The \emph{second dimension} regards how the resources used by the \glspl{cc} physically relate with the ones employed for the \gls{dc} used for data communication between the \gls{bs} and \glspl{ue}. We consider studying the following two options~\cite{bjornson2022reconfigurable}. First, an \emph{\gls{obcc}} uses communication resources that are orthogonal to the ones used by the \gls{dc}. More precisely, an \gls{obcc} exerts control over the propagation environment but is not affected by this control. Second, an \emph{\gls{ibcc}} uses the same communication resources as the \gls{dc}. This implies that the \gls{ibcc} reduces the available resources to transmit data, likely decreasing the performance of the overall system.} Furthermore, the successful transmission of the control messages toward the \gls{ris} depends on the current behavior of the \gls{ris} elements, meaning that the performance of the \gls{cc} is now dependent on how the \gls{ris} is configured.

\change{Our analysis suggests that employing either \gls{oce} or \gls{bsw} depends on the specific service requirements and the coherence time of the \gls{dc}.
If the scenario exhibits high channel coherence time, the \gls{oce} paradigm can potentially deliver superior data rate performance. However, its feasibility diminishes when the channel experiences rapid changes due to the increased control overhead. In contrast, the \gls{bsw} paradigm generally provides a lower data rate but proves suitable for scenarios with low coherence time due to its reduced control overhead. 
Additionally, our findings indicate that the reliability of control signaling is minimally affected by the paradigm used when the \gls{ris}-\gls{cc} is an \gls{obcc}. However, when an \gls{ibcc} serves as \gls{ris}-\gls{cc}, the results reveal that \gls{oce} control is less reliable than \gls{bsw} due to the increased complexity associated with the former in managing the \gls{ris}.}

The paper is organized as follows. Section~\ref{sec:model} presents the system model. 
Section~\ref{sec:paradigms} describes the transmission paradigms and their performance \emph{assuming error-free \glspl{cc}}. Section~\ref{sec:ris-control} shows how to analyze the impact of \glspl{cc} design in the communication performance, whose results are shown in Section~\ref{sec:results}.
Section~\ref{sec:extension} discusses how to relax some of the simplification assumptions made in this paper, while Section~\ref{sec:conclusions} concludes the paper.

\paragraph*{Notation}
Lower and upper case boldface letters denote vectors and matrices, respectively. \change{Calligraphic letters denote sets, whose cardinality is $|\cdot|$}. The Euclidean norm of $\mathbf{x}$ is $\lVert\mathbf{x}\rVert$; $\odot$ denotes the element-wise product. $\mc{P}(\cdot)$ is the probability of an event, $\mathbb{E}[\cdot]$ is the expected value; $\mc{CN}(\bm{\mu},\mb{R})$ is the complex Gaussian distribution with mean $\bm{\mu}$ and covariance matrix $\mb{R}$, $\mathrm{Exp}(\lambda)$ is the exponential distribution with mean value $1/\lambda$. $\lfloor a \rfloor$ is the nearest lower integer of $a$; $\mathbb{N}$ and $\mathbb{C}$ are the set of natural and complex numbers; $\Re\left\{\cdot\right\}$ returns the real part of a complex number, and $j\triangleq \sqrt{-1}$.

\section{System Model} \label{sec:model}
\begin{figure*}
    \centering
    \begin{subfigure}{0.4\textwidth}
        \centering
        \includegraphics[width=5cm]{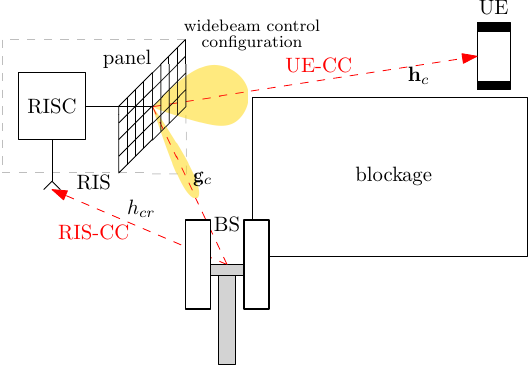}    
        \caption{Control flow.}
    \end{subfigure}  
    \begin{subfigure}{0.4\textwidth}
        \centering
        \includegraphics[width=5cm]{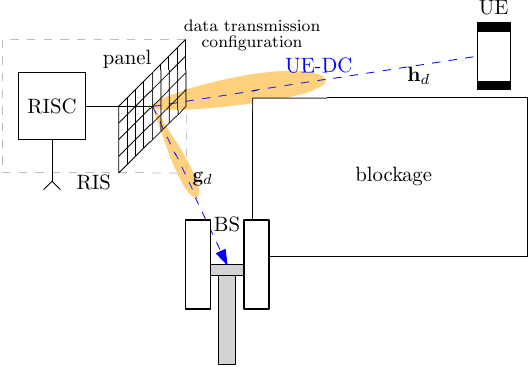}    
        \caption{Data flow.}
    \end{subfigure}
    \caption{Setup of interest: an \gls{ris} extends the coverage of the \gls{bs} which has a blocked link to the \gls{ue}. During control signaling, the \gls{risc} loads a \glsfirst{ctrl} configuration to the \gls{ris} to deliver low-rate control packets to the \gls{ue}. During data transmission, the \gls{bs} controls and communicates with the \gls{risc} to load a certain configuration to achieve a desired communication performance.}    
    \label{fig:system-model:scenario}
    \vspace{-0.5cm}
\end{figure*}

\change{
This section introduces a simplified communication model involving an \gls{ris}. We focus on the \gls{ul} scenario depicted in Fig.~\ref{fig:system-model:scenario}, which comprises a single-antenna \gls{bs}, a single-antenna \gls{ue}, and a solely reflective \gls{ris}~\cite{Tsinghua_RIS_Tutorial}. The primary aim of this system is to facilitate efficient data communication between the \gls{ue} and the \gls{bs} with the support of the \gls{ris}. To accomplish this, the \gls{bs} establishes control signaling links with both the \gls{ris} and the \gls{ue}. The details and the key assumptions are given below. 
}

\paragraph{RIS operation model}
\change{
The internal operations of the \gls{ris} are divided into the \gls{ris} panel and the \gls{risc}. The panel comprises $N$ elements equally spaced on a planar surface. The surface is solely reflective, implying that each element only controls the reflection properties of the incoming waves and that the \gls{ris} cannot process any of them. In particular, we focus on configuring the phases of the elements to change the reflection angle of an incoming wave, where their phase shifts are denoted as $\varphi_n \in [0,2\pi]$, $\forall n\in\mc{N}= \{1,\dots,N\}$\footnote{For the sake of simplicity and following the standard practice in literature, we consider an ideal \gls{ris} to show the theoretical performance achievable by the system at hand. We expect that more realistic models addressing attenuation, mutual coupling, quantization, and non-linear effects would reduce the overall performance~\cite{Tsinghua_RIS_Tutorial}.}. We denote as $\bm{\phi}= [e^{j\varphi_1}, \dots, e^{j\varphi_N}]\T\in \mathbb{C}^{N}$ the vector representing a particular \emph{configuration}, \emph{i.e.}, the set of phase shifts configured at the metasurface's elements at a given time. The \gls{risc} is in charge of loading different configurations to the \gls{ris} surface. Without loss of generality, we indicate as $\tau_s\in\mathbb{R}_+$ the time needed to switch to a new configuration. Moreover, the \gls{risc} is equipped with a communication module, which is used to exchange control signals with the \gls{bs}. We refer to the link connecting the \gls{risc} to the \gls{bs} as the \gls{ris}-\gls{cc}. We consider that the \gls{risc} stores a look-up table containing a set of predefined configurations, namely the \emph{common codebook} of configurations, $\mc{C}=\{\bm{\phi}_1,\dots,\bm{\phi}_C\}$, $|\mc{C}| = C$, that can be designed according to the task at hand; a copy of $\mc{C}$ is stored in the \gls{bs} as well. The control signals between the \gls{bs} and the \gls{risc} can then be based on indexing elements of $\mathcal{C}$ or by explicitly sending a configuration $\bm{\phi}$, if the \gls{bs} wants to load a configuration not present in the common codebook, \emph{i.e.}, $\bm{\phi}\notin\mc{C}$. In the next sections, $\mc{C}$ is going to be defined more explicitly according to \gls{oce} and \gls{bsw} paradigms.
}

\paragraph{UE communication model}\label{sec:system-model:ue-model}
\change{
To analyze the communication performance, we assume a frame-based fixed time system of $\tau\in\mathbb{R}_{+}$ seconds . The frame is further divided into phases that organize the signals exchanged at a given time. Within a frame, control and data information are exchanged in different phases. In each frame, the \gls{ue} communicate with the \gls{bs} through the \gls{ris}: it transmits payload data using the \gls{ue}-\gls{dc}, while the exchange of control messages uses the \gls{ue}-\gls{cc}. 
To ensure mathematical tractability, ideal timing and frequency synchronization among devices at the \gls{phy} layer is assumed. Perfect synchronization is also considered at the frame level across these devices. Also, we consider that the \gls{ris} can be configured at any time within a frame, and we make the following assumption about the behavior of the \gls{ris} when the \gls{bs} and the \gls{ue} exchange control information.
}

\begin{assumption}[\gls{ris} ctrl configuration]
\label{assu:ctrl}
   \change{To allow control signaling exchange between the \gls{bs} and the \gls{ue}, we consider that the \gls{ris} loads a \emph{\gls{ctrl} configuration} that ensures that control messages traveling through the \gls{ue}-\gls{cc} reach the destination\footnote{The design of the \gls{ctrl} configuration is out of the scope of this paper; potential candidates for \gls{ctrl} configurations are wide-width beams and hierarchical radiation patterns, which generally offer good reliability and coverage when low data rates are needed~\cite{RISBroadCoverage,alexandropoulos2022hierarchical}.}. Without loss of generality, we assume that the \gls{risc} loads the \gls{ctrl} configuration when in idle, \emph{i.e.}, anytime it does not receive any explicit command from the \gls{bs}.}
\end{assumption}


\paragraph{Channel models}
\change{We adopt the block fading model at a frame level, implying that the channel conditions remain constant throughout the frame duration. Remark that this binds the frame duration $\tau$ to the channel coherence time: the higher the coherence time, the longer the frame duration. The following assumption is made on the channels, further described in the sequel.}

\begin{assumption}[Narrowband channels]
\label{assu:narrowband}
   \change{To allow a simple analysis of both control and data channels, the three channels under consideration -- \emph{1}) the \gls{ue}-\gls{dc}, \emph{2}) the \gls{ue}-\gls{cc}, and \emph{3}) the \gls{ris}-\gls{cc} -- are considered to be narrowband.\footnote{The analysis can be straightforwardly extended to a wideband transmission, as discussed in Section~\ref{sec:extension}.}}
\end{assumption}

\subsubsection{UE-DC} 
This channel operates at a central frequency $f_d$ with a bandwidth of $B_d$. The \gls{ul} \gls{snr} can be calculated as: 
\begin{equation} \label{eq:snr:uedc}
    \gamma = \frac{\rho_u}{\sigma_b^2} |\bm{\phi}\T  (\mb{h}_d \odot \mb{g}_d)|^2 = \frac{\rho_u}{\sigma_b^2} |\bm{\phi}\T \mb{z}_d|^2,
\end{equation}
where $\mb{h}_d\in\mathbb{C}^N$ and $\mb{g}_d\in\mathbb{C}^N$ are the gains of \gls{ue}-\gls{ris} and \gls{ris}-\gls{bs} links, respectively. We further define the equivalent \gls{dc} as $\mb{z}_d = \mb{h}_d \odot \mb{g}_d \in \mathbb{C}^{N}$. The \gls{ue} transmit power is $\rho_u$, and $\sigma_b^2$ is the noise power at the \gls{bs} \gls{rf} chain\footnote{In the remainder of the paper, we assume that the \gls{bs} knows the transmit and noise powers denoted through this section, \emph{i.e.}, $\rho_u$, $\rho_b$, $\sigma_r^2$, $\sigma_u^2$ and $\sigma_b^2$: the transmit powers are usually determined by the protocol or set by the \gls{bs} itself; the noise powers can be considered static for a long time horizon and hence estimated previously through standard estimation techniques, \emph{e.g.},~\cite{Yucek2006noise}.}. \change{The configurations $\bm{\phi}$ supporting the \gls{ue}-\gls{dc} are subject to design and will be further specified in the following sections.}

\subsubsection{UE-CC} 
This channel operates at central frequency $f_u$ with a bandwidth of $B_u$. 
The \gls{ue}-\gls{cc} channel is defined as
\begin{equation} \label{eq:channel:uecc}
    h_{cu} = \bm{\phi}\T_\mathrm{ctrl} (\mb{h}_c \odot \mb{g}_c) = \bm{\phi}\T_\mathrm{ctrl} \mb{z}_c,
\end{equation}
where $\bm{\phi}_\mathrm{ctrl}$ is the \gls{ctrl} configuration (see Assumption~\ref{assu:ctrl}) and $\mb{h}_c\in\mathbb{C}^N$ and $\mb{g}_c\in\mathbb{C}^N$ are the gains of the \gls{ue}-\gls{ris} and \gls{ris}-\gls{bs} links, respectively. The equivalent end-to-end channel is $\mb{z}_c = \mb{h}_c \odot \mb{g}_c \in \mathbb{C}^N$. We consider a worst-case scenario where the channel~\eqref{eq:channel:uecc} has no \gls{los} component, \emph{i.e.}, it is distributed as $h_{cu} \sim \mc{CN}(0, \tilde{\lambda}_u)$; $\tilde{\lambda}_u$ is a term accounting for the large-scale fading -- known by the \gls{bs} -- which is dependent on the \gls{ctrl} configuration design. Hence, the instantaneous \gls{snr} measured at the \gls{ue} is
\begin{equation} \label{eq:snr:uecc}
    \Gamma_{u} = \frac{\rho_b}{\sigma_u^2} |h_{cu}|^2 \sim \text{Exp}\left(\frac{1}{\lambda_u}\right), 
\end{equation}
where $\lambda_u = \frac{\rho_b \tilde{\lambda}_u}{\sigma^2_u}$ denotes the average \gls{snr} at the \gls{ue}, being $\rho_b$ the \gls{bs} transmit power and $\sigma^2_u$ the \gls{ue}'s \gls{rf} chain noise power. 
\change{We make the following assumption on this channel.}
\begin{assumption} [\gls{ue}-\gls{cc} design]
    \label{assu:ue-cc}
    \change{
    We assume that the \gls{ue}-\gls{cc} operates as an \gls{ibcc}, meaning that the frequency $f_u$ matches the frequency $f_d$, and the physical resources allocated for the \gls{ue}-\gls{cc} coincide with those utilized for the \gls{ue}-\gls{dc}. This assumption is based on the premise that any \gls{ue}-\gls{cc} signal must travel through the \gls{ris} and the \gls{ris} operates at the same data frequency and bandwidth.
    }
\end{assumption}


\subsubsection{RIS-CC} 
This narrowband channel operates on central frequency $f_r$ with bandwidth $B_r$. Let $h_{cr}\in\mathbb{C}$ denote the channel coefficient of the \gls{ris}-\gls{cc}. To obtain simple analytical results, we assume that $h_{cr} \sim \mc{CN}(0, \tilde{\lambda}_r)$, where $\tilde{\lambda}_r$ accounts for the large-scale fading, assumed known by the \gls{bs}. Hence, the instantaneous \gls{snr} measured at the \gls{risc} is
\begin{equation} 
    \Gamma_{r} = \frac{\rho_b}{\sigma_r^2} |h_{cr}|^2 \sim \text{Exp}\left(\frac{1}{\lambda_r}\right),
    \label{eq:snr:riscc}
\end{equation}
where $\lambda_r = \frac{\rho_b \tilde{\lambda}_r}{\sigma_r^2}$ denotes the average \gls{snr} with $\sigma^2_r$ being the noise power at the \gls{risc} \gls{rf} chain. 
\change{We make the following assumption on the \gls{ris}-\gls{cc}.}
\begin{assumption}[\gls{ris}-\gls{cc} design]
    \label{assu:ris-cc}
    \change{The \gls{ris}-\gls{cc} can either be: $i$) \gls{ibcc}, implying that the physical resources employed by this channel are overlapping with the one used by the \gls{ue}-\gls{dc}, \emph{i.e.}, $f_r = f_d$;  or $ii$) \gls{obcc}, where the physical resources are orthogonal, \emph{e.g.}, simulating a wired connection between the \gls{bs} and the \gls{risc}. In the case of \gls{obcc}, we further assume that the \gls{ris}-\gls{cc} is an error-free channel, \emph{i.e.}, $\lambda_r\rightarrow\infty$, with feedback capabilities since the system designer can easily make the \gls{ris}-\gls{cc} as reliable as possible.}     
\end{assumption}




\section{RIS-Enabled Communication Paradigms} \label{sec:paradigms}
\change{In this section, we first describe the structure and building blocks of a generic \gls{ris}-enabled communication paradigm. We use this to instantiate two particular transmission strategies: the \gls{oce} and the \gls{bsw}, representing the multiplexing and diversity paradigms discussed in Section~\ref{sec:intro:contributions}}.
\change{We analyze their performance in terms of the expected \gls{snr} and \gls{se}, while specifying the errors eventually occurring during their operations.}

\subsection{Generic structure}
\change{In a system without \glspl{ris}, a generic structure for a typical transmission strategy can be divided into three key phases occurring in every frame: ``\emph{Signaling},'' ``\emph{Algorithmic},'' and ``\emph{Payload}.'' The Signaling phase encompasses the actions conducted on the \glspl{cc} required for network node control. This phase relies on the quality of the \glspl{cc} and the information content within the control messages. The Algorithmic phase involves operations aiming at optimizing transmission parameters, such as selecting the transmit \gls{se}. The specifics of this phase are contingent on the chosen communication paradigm. The Payload phase handles the actual data transmission.}

\change{In an \gls{ris}-aided system, we identify that a generic structure would have two Signaling phases, namely ``\emph{Initialization}'' and ``\emph{Setup},'' in conjunction with the ``\emph{Algorithmic}'' and ``\emph{Payload}'' phases. Sequentially, we have: Initialization, Algorithmic, Setup and Payload}\footnote{We note that there could be communication strategies in which some of these phases may not be present, \emph{e.g.}, access procedures; however, the mentioned four phases set a basis for a sufficiently general framework for \gls{ris}-aided communication that can be used, in principle, to design other schemes where some of the steps are merged or omitted, as discussed in Section~\ref{sec:extension}.}. \change{The time of a frame is thus divided as $\tau=\tau\ini+\tau\alg+\tau\set+\tau\pay$, where $\tau_i < \tau$, with $i\in\{{\rm ini},{\rm alg},{\rm set},{\rm pay}\}$, is the time of the corresponding phase. The duration of each phase can vary depending on the paradigm and the kind of \glspl{cc}.} The phases are further elaborated as follows.


\paragraph{Initialization} 
The \gls{bs} notifies the \gls{ris} and \gls{ue} about the beginning of a frame over the \glspl{cc}. It is assumed that the \gls{risc} loads the \gls{ctrl} configuration at the start of this phase. 
\change{Although not considered here, the Initialization phase can also incorporate a random access procedure (see, \emph{e.g.},~\cite{Croisfelt2023}) as an intermediate step where newly connected \glspl{ue} are scheduled.}

\paragraph{Algorithmic} 
\change{
This phase encompasses all the processes and computations needed to optimize the subsequent \change{Payload phase}, where the actual data transmission takes place. 
Objectives of this phase encompass the selection/optimization of the appropriate \gls{ris} configuration(s), while others, such as determining the transmission parameters for the \gls{ue} and/or the \gls{bs}, could be included. To tackle these objectives, some form of wireless environment sensing from the end nodes is required, typically enabled by the transmission of pilot sequences, whose specifications -- their number, waveform, \gls{ul} or \gls{dl} transmission, etc -- are defined by the transmission strategy. The \gls{bs} can then use the collected pilot signals and invoke pre-defined algorithms to fulfil the objectives. The outcome of these algorithms might be affected by different types of \emph{algorithmic errors} that may prevent the system from performing as expected, and thus, should be considered when analyzing the overall performance.
}

\paragraph{Setup}
\change{
During this phase, the \gls{ris} configuration chosen during the Algorithmic phase needs to be communicated to the \gls{risc}, which in turn commands the \gls{ris} to load it. Additionally, further control signaling may occur between the \gls{bs} and the \gls{ue} as a final check before data transmission to, e.g., set the \gls{mcs}.} 

\paragraph{Payload}
\change{Here, the actual data transmission takes place while the \gls{ris} loads the configuration specified before. This phase may or may not include feedback of the data sent by the \gls{ue} at the end; this aspect is not considered in this paper. The communication performance of the considered \gls{ris}-enabled communication system is measured during this phase.}

\begin{figure*}[!t]
    \centering
    \begin{subfigure}[t]{0.45\linewidth}
        \centering
        \includegraphics[width=0.9\textwidth]{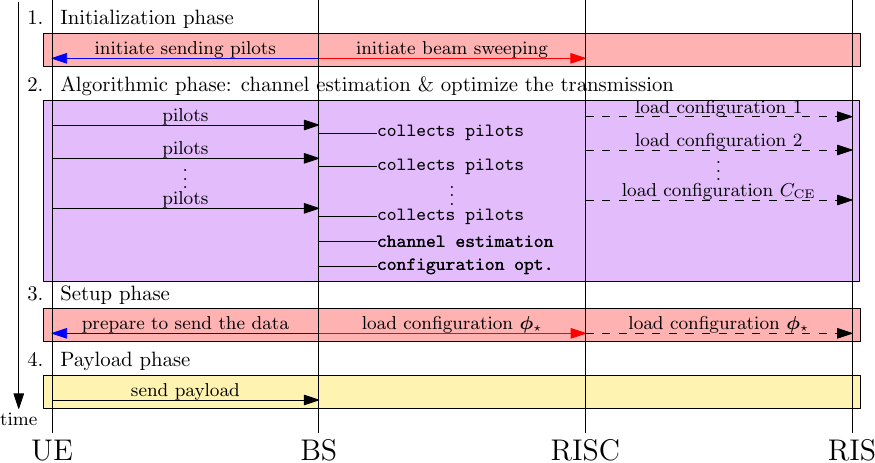}
        \caption{\gls{oce} paradigm.}
        \label{fig:RIS-oce}
    \end{subfigure}
    \hfill
    \begin{subfigure}[t]{0.45\linewidth}
        \centering
        \includegraphics[width=0.9\textwidth]{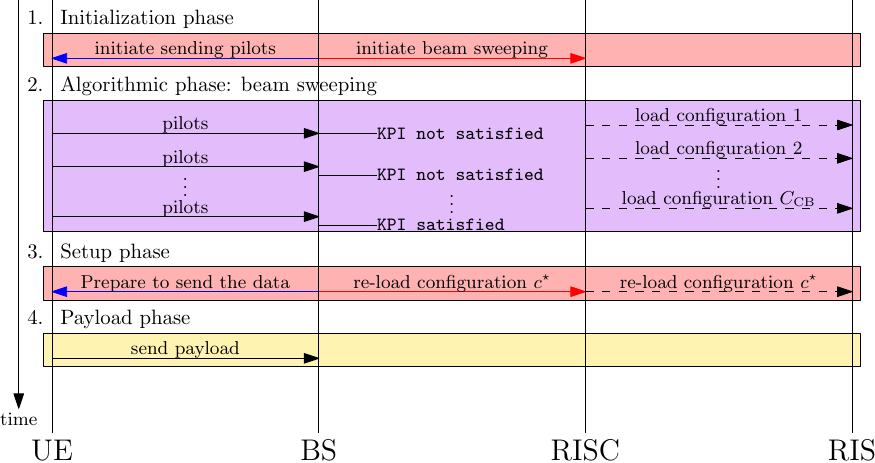}
        \caption{\gls{bsw} paradigm.}
        \label{fig:RIS-bsw}
    \end{subfigure} 
    \caption{Data exchange diagram of the two \gls{ris}-enabled communication paradigms. Signals traveling through \gls{ris}-\gls{cc}, \gls{ue}-\gls{cc}, and \gls{ue}-\gls{dc} are represented by solid \textcolor{red}{red}, solid \textcolor{blue}{blue}, and solid black lines, respectively. \gls{risc} to \gls{ris} commands are indicated with dashed black lines. \gls{bs} operations are in \texttt{monospaced font}.}
    \label{fig:RIS-frames}
\end{figure*}

\change{In the following subsections, we describe two state-of-the-art paradigms using the generic structure defined above. After a general description, we investigate their Algorithmic phases, analyzing their performance and possible errors.} 
The first paradigm is the \emph{\gls{oce}}, which follows a multiplexing transmission: the \gls{ue}'s \gls{csi} is evaluated at the \gls{bs} and then exploited to compute the \gls{ris}' optimal configuration and the corresponding achievable data rate. \change{Then, the transmission is made using the optimized configuration and \gls{mcs}.} 
The second approach is the \emph{\gls{bsw}}, defined as a communication paradigm in~\cite{An2022}, but already used in previous works (\emph{e.g.},~\cite{Jamali2022, alexandropoulos2022hierarchical, Croisfelt2022randomaccess}). \change{This paradigm resembles the concept of diversity transmission: the \gls{bs} selects a target \gls{kpi} \emph{a priori}; then, it instructs the \gls{risc} to sweep through a set of predefined configurations, expecting that at least one will satisfy the target \gls{kpi}; then, the transmission is made using the chosen configuration.}
Fig.~\ref{fig:RIS-frames} shows the data exchange diagrams of the two paradigms, comprised of \gls{cc} messages, configuration loading, processing operations, and data transmission. 

\subsection{Optimization based on Channel Estimation (OPT-CE)}\label{sec:communication-paradigms:oce}
In \gls{oce}, the \gls{bs} needs to obtain the \gls{csi} for the \gls{ue} to optimize the \gls{ris} configuration. The necessary measurements can be collected by transmitting pilot sequences from the \gls{ue}. During the Initialization phase, the \gls{bs} informs the other entities that the procedure is starting. First, the \gls{ue} is informed through the \gls{ue}-\gls{cc} to prepare to send pilots. Second, to solve the indeterminacy of the $N$-path \gls{ce} due to the \gls{ris} presence~\cite{Wang2020}, the \gls{ris} is instructed to sweep through a common codebook of configurations during the Algorithmic phase, called \emph{\gls{ce} codebook} and denoted as $\mc{C}\oce \subseteq \mc{C}$. To change between the configurations in the \gls{ce} codebook, we consider that the \gls{bs} needs to send only a single control message to the \gls{risc} since the \gls{ris} sweeps following the order stipulated by the codebook. During the Algorithmic phase, the \gls{ue} sends replicas of its pilot sequence, subject to different \gls{ris} configurations, to let each of them experience a different propagation environment. After a sufficient number of samples is received, the \gls{bs} estimates the \gls{csi} and compute the optimal \gls{ris}'s configuration~\cite{Tsinghua_RIS_Tutorial}. Then, the Setup phase starts, in which the \gls{bs} uses the the ctrl configuration\footnote{We remark that the \gls{ctrl} configuration is automatically loaded after the Algorithmic phase ends due to the idle state of the \gls{ris}. Another approach is loading the optimal configuration evaluated in the Algorithmic phase also to send the control information toward the \gls{ue}; nevertheless, the \gls{risc} needs to be informed previously by a specific control message by the \gls{bs}. We do not consider this approach to keep the frame structure of the two paradigms similar, thereby simplifying the analysis and the presentation in Section~\ref{sec:ris-control}.} to implement the \gls{ue}-\gls{cc} and instruct the \gls{ue} to start sending data. Subsequently, the \gls{bs} informs the \gls{risc}, over the \gls{ris}-\gls{cc}, to load the optimal configuration. Finally, the Payload phase takes place.

\subsubsection{Performance Analysis}
We now present the \gls{ce} procedure and analyze its performance in connection to the cardinality of the employed codebook $C\oce = |\mc{C}\oce|$. The considered method employed is a simplified version proposed in~\cite{Wang2020}. Let us start with the pilot sequence transmission and its processing. 
We denote a single pilot sequence as $\bm{\psi} \in \mathbb{C}^{p}$, spanning $p$ symbols and having $\lVert \bm{\psi} \rVert^2 = p$. Every time a configuration from the codebook is loaded at the \gls{ris}, the \gls{ue} sends a replica of the sequence $\bm{\psi}$ towards the \gls{bs}. When the configuration $c\in\mc{C}\oce $ is loaded, the following signal is received at the \gls{bs}
\begin{equation} \label{eq:oce:replica}
    \mb{y}_c\T  = \sqrt{\rho_u} \bm{\phi}_c\T \mb{z}_d  \, \bm{\psi}\T + \tilde{\mb{w}}_c\T \in\mathbb{C}^{1 \times p},
\end{equation}
where $\bm{\phi}_c$ denotes the phase profile vector of the configuration $c\in\mc{C}\oce$, $\rho_u$ is the transmit power, and $\tilde{\mb{w}}_c\sim \mc{CN}(0, \sigma_b^2 \mb{I}_p)$ is the \gls{awgn}. The received symbol is then correlated with the pilot sequence and normalized by $\sqrt{\rho_u} p$, yielding
\begin{equation} \label{eq:oce:pilotprocess}
    y_c  = \frac{1}{\sqrt{\rho_u} p}\mb{y}_c\T \bm{\psi}^* = \bm{\phi}_c\T \, \mb{z}_d + w_c \in\mathbb{C},
\end{equation}
where $w_c\sim\mc{CN}(0, \frac{\sigma_b^2}{p \rho_u})$ is the resulting \gls{awgn}.\footnote{The consideration of dividing the pilot transmission over configurations over small blocks of $p$ symbols serves three purposes: $i$) from the hardware point of view, it might be difficult to change the phase-shift profile of an \gls{ris} within the symbol time, $ii$) to reduce the impact of the noise, and $iii$) to have the possibility of separating up to $p$ \gls{ue}'s data streams, if the pilots are designed to be orthogonal to each other~\cite{massivemimobook}.}
The pilot transmissions and the processing in~\eqref{eq:oce:pilotprocess} are repeated for all \gls{ris} configurations, \emph{i.e.}, $\forall c\in\mc{C}\oce$. The resulting signal $\mb{y} = [y_1, y_2, \dots, y_{C\oce}]\T\in\mathbb{C}^{C\oce}$ can be then compactly written in the following form:
\begin{equation} \label{eq:oce:signal}
    \mb{y}= \mb{\Theta}\T \mb{z}_d + \mb{w},
\end{equation}
where $\bm{\Theta} = [\bm{\phi}_1, \bm{\phi}_2, \dots, \bm{\phi}_{C\oce}]\in\mathbb{C}^{N\times C\oce }$ is the matrix containing all the configurations used and $\mb{w} = [w_1, \dots, w_{C\oce}]\T \in\mathbb{C}^{C\oce}$ is the \gls{awgn} term. For the sake of generality, we will assume that there is no prior information about the channel distribution at the \gls{bs}. Therefore, we cannot estimate separately $\mb{h}_d$ and $\mb{g}_d$, but only the cascade channel $\mb{z}_d$. 
It is possible to show that a necessary (but not sufficient) condition to perfectly estimate the channel coefficients is that $C\oce\geq{N}$~\cite{Wang2020}. Indeed, we want a linearly independent set of equations, which can be obtained by constructing the configuration codebook for \gls{ce} to be at least of rank $N$. As an example, we can use the \gls{dft} matrix, \emph{i.e.}, $[\bm{\Theta}]_{n,c} = e^{-j2\pi \frac{(n-1) (c-1)}{C\oce}}$, with $n=\mc{N}$ and $c\in\mc{C}\oce$,
with $\bm{\Theta}^* \bm{\Theta}\T = C\oce \mb{I}_{N}$. Considering that the parameter vector of interest is deterministic, the least-squares estimate yields the estimation~\cite{Kay1997}
\begin{equation}
    \hat{\mb{z}}_d=\dfrac{1}{C\oce} \bm{\Theta}^* \mb{y} = \mb{z}_d + \mb{n},
\end{equation}
where $\mb{n}\sim\mc{CN}(0, \frac{\sigma_b^2}{p \rho_u C\oce } \mb{I}_N)$, and whose performance is proportional to $1/C\oce$.
%
Based on the estimated equivalent channel, the \gls{bs} can obtain the configuration $\bm{\phi}_\star$ that maximizes the instantaneous \gls{snr} of the typical \gls{ue}, as follows: 
\begin{equation}
    \begin{aligned}
        \bm{\phi}_\star &= \max_{\phi}\left \{ \lvert \bm{\phi}\T \hat{\mb{z}}_d \rvert^2 \, \big| \, \lVert\bm{\phi}\rVert^2 = N \right\},
    \end{aligned}
\end{equation}
which turns out to provide the intuitive setting $(\bm{\phi}_\star)_n=e^{-j\angle{(\hat{\mb{z}}_d)_n}}$, $\forall n \in \mc{N}$. The \gls{ul} estimated \gls{snr} at the \gls{bs} is:
\begin{equation}
    \hat{\gamma}\oce = \frac{\rho_u}{\sigma_b^2} |\bm{\phi}_\star\T \hat{\mb{z}}_d|^2.
\end{equation}
Based on the estimated \gls{snr}, the \gls{se} of the data communication can be adapted to be the maximum achievable, \emph{i.e.},
\begin{equation} \label{eq:oce:se}
    \eta\oce = \log_2(1 + \hat{\gamma}\oce).
\end{equation}

\subsubsection{Algorithmic errors}
For the \gls{oce} paradigm, a communication outage occurs in the case of an \emph{overestimation error}, \emph{i.e.}, if the selected \gls{se} $\eta\oce$ is higher than the actual channel capacity, leading to an unachievable communication rate~\cite{Shannon1949}. The probability of this event is
\begin{equation} \label{eq:oce:ae1}
    p_\mathrm{ae} = \mc{P} \left[\eta\oce = \log(1 + \hat{\gamma}\oce) \ge \log_2\left(1+ \gamma\oce \right) \right],
\end{equation}
where $\gamma\oce = \frac{\rho_u}{\sigma_b^2} |\bm{\phi}_\star\T \mb{z}_d|^2$ is the actual \gls{snr} at the \gls{bs}. Eq.~\eqref{eq:oce:ae1} translates to the condition
\begin{equation} \label{eq:oce:ae2}
    p_\mathrm{ae} = \mc{P}\left[\hat{\gamma}\oce \ge \gamma\oce\right] = \mc{P}\left[ |\bm{\phi}_\star\T \mb{z}_d + \bm{\phi}_\star\T \mb{n}|^2 \ge  |\bm{\phi}_\star\T \mb{z}_d|^2 \right].
\end{equation}
A detailed analysis of~\eqref{eq:oce:ae2} relies on the channel model of $\mb{z}_d$, and thereby on a prior assumption about the scenario (\emph{e.g.},~\cite{RIS_Nakagami}); we therefore numerically evaluate  the impact of the \gls{oce} algorithmic error.


\subsection{Codebook-Based Beam Sweeping (CB-BSW)}\label{sec:communication-paradigms:bsw}
In \gls{bsw}, the \gls{bs} now does not require explicit \gls{csi} of the \gls{ue}. In the Initialization phase, the \gls{bs} commands the start of a new frame by signaling to the \gls{ris} and \gls{ue}. \change{A \emph{\gls{BSW} process}, \emph{i.e.}, an \gls{ris} configuration selection, is performed during the Algorithmic phase. This process comprises the \gls{ue} sending reference signals, while the \gls{bs} commands the \gls{ris} to change its configuration at regular periods accordingly to a set of predefined configurations, labeled as the \emph{\gls{BSW} codebook} denoted by $\mc{C}\bsw \subseteq \mc{C}$.} The \gls{bs} receives the reference signals that are used to measure \gls{ue}'s \gls{kpi}. The \gls{BSW} process is triggered when a single \gls{bs} control message is received by the \gls{risc}. At its end, the \gls{bs} selects a configuration satisfying the target \gls{kpi}. During the Setup phase, the \gls{bs} informs the \gls{ue} over the \gls{ue}-\gls{cc} to prepare to send data, while the \gls{ris} uses the \gls{ctrl} configuration, and informs the \gls{risc} through the \gls{ris}-\gls{cc} to load the selected configuration. Finally, the Payload phase takes place. 

\begin{remark}[Fixed vs Flexible frames]
    \change{The \gls{BSW} process during the Algorithmic phase may make use of i) a \emph{fixed} or ii) a \emph{flexible} frame structure. The fixed frame ends after a fixed number of \gls{BSW} codebook configurations have been loaded. The flexible frame structure allows stopping the \gls{BSW} as soon as a \gls{kpi} value measured is above the target one. Flexible frame requires on-the-fly \gls{kpi} measurements at the the \gls{bs}, while \gls{ue}-\gls{cc} needs to be reserved to promptly inform the \gls{ue} about the frame termination when the target \gls{kpi} is met, thus modifying the overall frame (see Section~\ref{sec:ris-control}).}
    \label{remark:cb-bsw:fixed-vs-flexible}
\end{remark}

\subsubsection{Performance analysis}
For \gls{bsw} it is necessary to assume that the target \gls{kpi} is a target \gls{snr} $\gamma_0$ measured at the \gls{bs} via the average \gls{rss} metric. In this case, a fixed \gls{se} is considered \emph{a priori}, which is given by
\begin{equation} \label{eq:bsw:se}
    \eta\bsw = \log_2(1 + \gamma_0),
\end{equation}
and the goal is find a configuration from the \gls{ris} codebook that supports it. We analyze the system performance starting from the pilot sequence transmission and processing. As before, every pilot sequence consists of $p$ symbols\footnote{The pilot sequences for \gls{oce} and \gls{bsw} can be different and have different lengths. In practice, they should be designed and optimized for each of those approaches, which is beyond the scope of this paper. We use the same pilot sequence length notation in both paradigms for simplicity.}. Once again, we denote a single sequence as $\bm{\psi} \in \mathbb{C}^{p}$ having $\lVert\bm{\psi} \rVert^2 = p$. After the \gls{ris} loads configuration $c\in\mc{C}\bsw$, the \gls{ue} sends a replica of $\bm{\psi}$ and, similar to~\eqref{eq:oce:replica}, the \gls{bs} receives the signal: 
\begin{equation} \label{eq:bsw:replica}
    \mb{y}_c\T  = \sqrt{\rho_u} \bm{\phi}_c\T \mb{z}_d \bm{\psi}\T + \tilde{\mb{w}}_c\T \in\mathbb{C}^{1 \times p},
\end{equation}
where $\bm{\phi}_c$ denotes the configuration $c \in \mc{C}\bsw$. The received signal is then correlated with $\bm{\psi}$ and normalized by $p$:
\begin{equation} \label{eq:bsw:pilotprocess}
    y_c  = \frac{1}{p}\mb{y}_c\T \bm{\psi}^* = \sqrt{\rho_u} \bm{\phi}_c\T \mb{z}_d + w_c \in\mathbb{C},
\end{equation}
where $w_c\sim\mc{CN}(0, \frac{\sigma_b^2}{p})$ is the resulting \gls{awgn}. 
The \gls{snr} provided by the configuration can be estimated as follows: 
\begin{equation} \label{eq:bsw:gammahat}
    \hat{\gamma}_c = \frac{|y_c|^2}{\sigma_b^2} \hspace{-0.3mm} = \hspace{-0.3mm} \underbrace{\frac{\rho_u}{\sigma_b^2} |\bm{\phi}_c\T \mb{z}_d|^2 }_{\gamma_c} \hspace{-0.1mm}+\hspace{-0.2mm} 2 \Re\hspace{-0.5mm}\left\{ \frac{\sqrt{\rho_u}}{\sigma_b^2} \bm{\phi}_c\T \mb{z}_d\, w_c\right\} \hspace{-0.6mm}+\hspace{-0.5mm} \frac{|w_c|^2}{\sigma_b^2},
\end{equation}
where $|w_c|^2 \sigma_b^{-2} \sim \mathrm{Exp}(p)$. 
It is worth noting that the estimated \gls{snr} is affected by the exponential error generated by the noise, but also by the error of the mixed product between the signal and the noise, whose \gls{pdf} depends on the \gls{pdf} of $\mb{z}_d$. Based on~\eqref{eq:bsw:gammahat}, we can select the best configuration $c^\star\in\mc{C\bsw}$ providing the target \gls{kpi}. According to Remark~\ref{remark:cb-bsw:fixed-vs-flexible}, we next discuss the selection of the configuration for the two different frame structures.

\paragraph{Fixed Frame} 
When the frame has a fixed structure, the \gls{BSW} procedure ends after the \gls{ris} sweeps through the whole codebook. In this case, we can measure the \glspl{kpi} for all available configurations. The configuration selected for the payload phase is the one achieving the highest estimated \gls{snr} among the ones satisfying the target \gls{kpi} $\gamma_0$, as
\begin{equation} \label{eq:bsw:cstar:fixed}
    c^\star = \argmax_{c\in\mc{C}\bsw} \{\hat{\gamma}_c \,|\, \hat{\gamma}_c \ge \gamma_0\}.
\end{equation}
If no configuration achieves the target \gls{kpi}, the communication is not feasible, and we run into an outage event.

\paragraph{Flexible Frame} 
When the frame has a flexible structure, the end of the \gls{BSW} process is triggered by the \gls{bs} when the measured \gls{kpi} reaches the target value. A simple on-the-fly selection method involves testing if the estimated \gls{snr} is greater than the target $\gamma_0$; \emph{i.e.}, after eq.~\eqref{eq:bsw:gammahat} is obtained for configuration $c\in\mc{C}\bsw$, we set
\begin{equation} \label{eq:bsw:cstar:flexible}
    c^\star = c \iff \hat{\gamma}_c \ge \gamma_0.
\end{equation}
As soon as $c^\star$ is found, the \gls{bs} communicates to both \gls{ris} and \gls{ue} that the Payload phase can start; otherwise, the \gls{BSW} process continues until a configuration is selected. If no configuration of the codebook $C\bsw$ satisfies the condition~\eqref{eq:bsw:cstar:flexible}, then communication is not feasible and outage occurs.

\subsubsection{Algorithmic errors}
For the \gls{bsw} paradigm, an outage event occurs when no configuration in the \gls{BSW} codebook satisfies the target \gls{kpi}, and when the selected configuration provides an \gls{snr} lower than the target one, although the estimated \gls{snr} was higher; we denote the latter as the overestimation event. These two events are mutually exclusive, and hence, their probability is
\begin{equation} \label{eq:bsw:ae}
\begin{aligned}
    p_\mathrm{ae} &= \mc{P} \left[ \gamma_{c^\star} \le \gamma_0 | \hat{\gamma}_{c^\star} > \gamma_0 \right] + \mc{P}\left[ \hat{\gamma}_{c} \le \gamma_0, \, \forall c\in\mc{C}\bsw \right]  \\
    &= \mc{P} \left[ \hat{\gamma}_{c^\star} - \gamma_0 \le  \frac{|w_{c^\star}|^2}{\sigma_b^2} + 2 \Re\left\{ \frac{\sqrt{\rho_u}}{\sigma_b^2} \bm{\phi}_{c^\star}\T \mb{z}_d\, w_c\right\} \right] \\ &\quad+ \mc{P}\left[ \hat{\gamma}_{1} \le \gamma_0, \dots, \hat{\gamma}_{C\bsw} \le \gamma_0  \right],
\end{aligned}    
\end{equation}
where $\gamma_{c^\star} = \frac{\rho_u}{\sigma_b^2} | \bm{\phi}_{c^\star}\T \mb{z}_d|^2$ is the actual \gls{snr} and $\hat{\gamma}_{c^\star} - \gamma_0  > 0$.
By applying Chebychev inequality, the overestimation probability (first term) can be upper bounded by
    \begin{equation} \label{eq:bsw:oebound}
     \mc{P} \left[ \hat{\gamma}_{c^\star} \hspace{-1mm} - \hspace{-0.8mm} \gamma_0 \hspace{-.5mm}\le \hspace{-0.8mm} \frac{|w_{c^\star}|^2}{\sigma_b^2} \hspace{-.5mm}+\hspace{-.5mm} 2 \Re\hspace{-.5mm}\left\{ \frac{\sqrt{\rho_u}}{\sigma_b^2} \bm{\phi}_{c^\star}\T \mb{z}_d\, w_c\hspace{-.5mm}\right\} \right] 
     \hspace{-1mm}\le \hspace{-1mm}\frac{p^{-1}}{\hat{\gamma}_{c^\star} \hspace{-1mm}-\hspace{-.8mm} \gamma_0},
\end{equation}
from which we infer that the higher the gap between $\hat{\gamma}_{c^\star}$ and $\gamma_0$, the lower the probability of error. The \gls{bsw} employing the fixed structure generally has a higher value of ${\hat{\gamma}_{c^\star} - \gamma_0}$ than the one with the flexible structure due to the use of the $\argmax$ operator to select the configuration $c^\star$. Therefore, the fixed structure is generally more robust to overestimation errors. 
On the other hand, the evaluation of the probability that no configuration in the beam sweeping codebook satisfies the target \gls{kpi} requires the knowledge of the \gls{cdf} of the estimated \gls{snr}, whose analytical expression is channel-model dependent and generally hard to obtain.
Here, we also resort to numerical evaluation of the impact of the \gls{bsw} algorithmic errors.

\subsection{Trade-offs in different transmission paradigms}
\label{sec:paradigms:comment}
The two aforedescribed \gls{ris}-aided transmission paradigms can be seen as a generalization of the \emph{fixed rate} (multiplexing) and \emph{adaptive rate} (diversity) transmission approaches. Essentially, the \gls{se} of the \gls{oce} is adapted to the achievable rate under the optimal configuration (see~\eqref{eq:oce:se}), while the \gls{se} of the \gls{bsw} is set \emph{a priori} according to the target \gls{kpi} (see~\eqref{eq:bsw:se}). Comparing~\eqref{eq:oce:se} and~\eqref{eq:bsw:se} under the same environmental conditions, we have that 
\begin{equation}
    \eta\bsw \le \eta\oce,
\end{equation}
where the price to pay for the higher \gls{se} of the \gls{oce} paradigm is the increased overhead. \gls{oce} needs an accurate \gls{csi} for reliable rate adaptation, which translates into a higher number of sequences to be transmitted by the \gls{ue} compared to \gls{bsw}. Furthermore, additional time and processing are required to determine the optimal configuration of the \gls{ris}. Consequently, the \gls{se} of data transmission alone cannot be considered a fair comparison metric, as it does not consider the overheads generated by the communication paradigms. 

\section{Impact of the Control Channels} 
\label{sec:ris-control}
In this section, we define a performance metric that simultaneously measures the communication performance and the impact of control signaling. We then characterize the terms of this metric regarding the overhead and the reliability of the signaling for the presented paradigms. 

\subsection{Utility function definition}
To measure the communication performance, we define a utility function that takes into account \emph{a}) the overhead and the error of the communication paradigms and \emph{b}) the reliability of the \glspl{cc}. Regarding the former, \change{we define the \emph{goodput} $R$ as a discrete random variable whose value depends on the communication paradigm and its algorithmic errors:}
\begin{equation} \label{eq:netthroughput}
    R(\tau\pay, \eta) = 
    \begin{cases}
        \frac{\tau\pay}{\tau} B_d \, \eta, \text{ with prob. } 1 - p_\mathrm{ae},\\
        0, \text{ with prob. } p_\mathrm{ae},
    \end{cases}
\end{equation}
In this expression, $\eta = \eta\oce$ in~\eqref{eq:oce:se} or $\eta = \eta\bsw$ in~\eqref{eq:bsw:se} if \gls{oce} or \gls{bsw} is respectively employed, $\tau\pay$ is the duration of the payload phase, and $\tau$ is the overall frame duration. The overall overhead time is the sum of the time to carry out the Initialization, Algorithmic, and Setup phases, denoted as $\tau\ini$, $\tau\alg$, and $\tau\set$, respectively\footnote{\change{We remark that the overhead time directly impacts on the latency experienced by the \gls{ue}: given a fixed frame duration, a higher overhead translates into a lower time opportunity for the Payload phase, reducing the transmitted data in each slot, and hence, increasing the overall latency.}}. Accordingly, the payload time can be written as $\tau\pay = \tau - \tau\ini - \tau\alg - \tau\set$.
While the overall frame length is fixed, the overhead time depends on the transmission paradigm, being a function of: the duration of a pilot, $\tau_p$, and the number of replicas transmitted; the optimization time, $\tau_A$; and the time to control the \gls{ris}, composed of the time employed for the transmission of the control packets to the \gls{ue} (\gls{risc}), $\tau_{i}^{(u)}$ ($\tau_{i}^{(r)}$), and the time needed by the \gls{ris} to switch configuration, $\tau_s$.

Regarding the reliability of the \glspl{cc}, we denote as $P = P_u + P_r$ the total number of control packets needed to let a communication paradigm work, where $P_u$ and $P_r$ are the numbers of control packets intended for the \gls{ue} and the \gls{risc}, respectively. Whenever one of such packets is erroneously decoded or lost, an event of \emph{erroneous control} occurs. We assume that these events are independent of each other (and of the algorithmic errors). We denote the probability of erroneous control on the packet $i$ toward entity $k\in\{u,r\}$ as $p_i^{(k)}$, with $i\in\{1, \dots, P_k\}$ and $k \in\{u,r\}$. Erroneous controls may influence the overhead time and the communication performance: \change{the \gls{ris} configuration might change unpredictably}\footnote{\change{In our scenario, if the control packet is not received, the \gls{risc} will load the \gls{ctrl} configuration, \emph{i.e.}, a predictable configuration change. However, if the \gls{risc} receives, but incorrectly decodes, a control packet, the \gls{bs} cannot know which configuration, if any, will be loaded.}}, leading to a degradation of the performance, or worse, letting the data transmission fail. While losing a single control packet may be tolerable depending on its content, we assume all control packets must be received correctly to make the communication successful. In other words, no erroneous control event is allowed. Consequently, the probability of correct control is
\begin{equation} \label{eq:pcc}
    p_\mathrm{cc} = \prod_{k\in\{u,r\}} \prod_{i=1}^{P_k} \left(1 - p_i^{(k)}\right).
\end{equation}

\change{We can include the control reliability in the communication performance, taking into account the probability of correct control in the goodput metric in~\eqref{eq:netthroughput}. By assuming that the control and algorithmic errors are independent, the goodput is re-expressed as follows:}
\begin{equation} \label{eq:netthroughput:2}
    R(\tau\pay, \eta) = 
    \begin{cases}
        \frac{\tau\pay}{\tau} B_d \, \eta, \text{ with prob. } p_\mathrm{cc}(1 - p_\mathrm{ae}),\\
        0, \text{ with prob. } 1 - p_\mathrm{cc} (1 - p_\mathrm{ae}),
    \end{cases}
    \end{equation}
\change{Hence, the performance of the considered \gls{ris}-enabled communication system can be described by averaging $R$ w.r.t. the control, obtaining the following \emph{utility function}:}
\begin{equation} \label{eq:utility}
    U(\tau\pay, \eta) \hspace{-0.5mm}
    = \hspace{-0.5mm} p_\mathrm{cc} (1 - p_\mathrm{ae}) \hspace{-1mm}\left( 1 - \frac{\tau\ini + \tau\alg + \tau\set}{\tau}\right) B_d  \eta.
\end{equation}

\subsection{Overhead evaluation} \label{sec:overhead}

\begin{figure}[tbh]
    \centering
    \begin{subfigure}{\columnwidth}
        \centering
        \includegraphics[width=\textwidth]{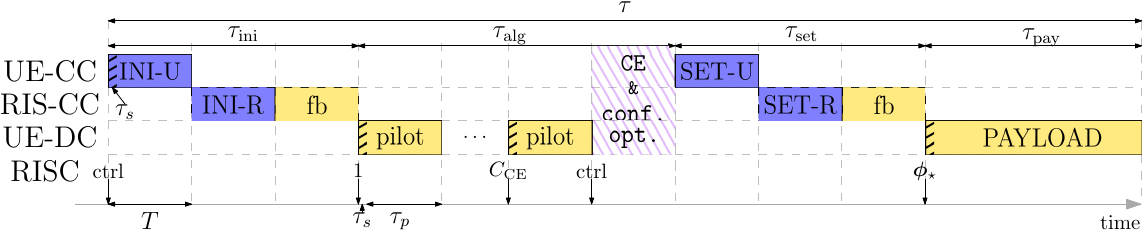}        
        \caption{\gls{oce}}
    \end{subfigure}
    \begin{subfigure}{\columnwidth}
        \centering
        \includegraphics[width=\textwidth]{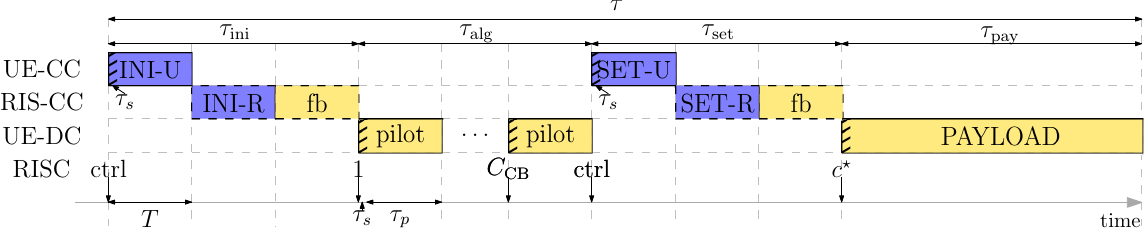}
        \caption{\gls{bsw}: fixed frame structure}
    \end{subfigure}
    \begin{subfigure}{\columnwidth}
        \centering
        \includegraphics[width=\textwidth]{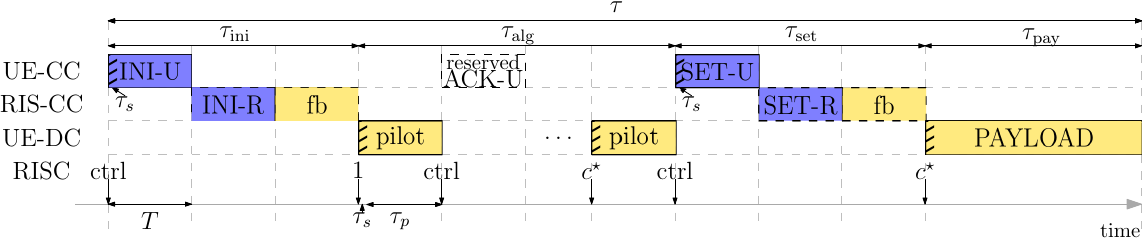}
        \caption{\gls{bsw}: flexible frame structure}
    \end{subfigure}       
    \caption{Frame structure for the communication paradigms under study. Packets colored in \textcolor{blue}{blue} and in \textcolor{yellow}{yellow} have \gls{dl} and \gls{ul} directions, respectively. Remark that INI-R (SET-R) packet and its feedback (fb) are sent at the same time as the INI-U (SET-U), but on different resources, if \gls{obcc} is present (dashed lines).}
    \label{fig:data-frames}
\end{figure}

Following the description of Section~\ref{sec:paradigms}, we present in Fig.~\ref{fig:data-frames} the frame structures of the two considered communication paradigms used to evaluate the induced overhead, where the rows represent the time horizon of the packets traveling on the different channels (first three rows) and the configuration loading time at the \gls{risc} (last row).
The time horizon is obtained assuming that all the operations span multiple numbers of \glspl{tti}, each of duration of $T$ seconds with $\lceil\tau / T\rceil \in \mathbb{N}$ being the total number of \glspl{tti} in a frame. At the beginning of each \gls{tti}, if the \gls{risc} loads a new configuration, the first $\tau_s$ seconds of data might be lost due to the unpredictable response of the channel during this switching period. When needed, we consider a guard period of $\tau_s$ seconds in the overhead evaluation to avoid data disruption. Remember that the \gls{risc} loads the \gls{ctrl} configuration any time it is in an idle state, \emph{i.e.}, at the beginning of the Initialization and Setup phases.

In Fig.~\ref{fig:data-frames}, we note that the overhead generated by the Initialization and Setup phases is \emph{transmission paradigm independent}\footnote{The reliability is still dependent on the paradigm (see Section~\ref{sec:reliability}).}, while it is \emph{\gls{cc} dependent}.  Both paradigms make use of $P = 4$ control packets, $P_u = 2$ control packets sent on the \gls{ue}-\gls{cc} and $P_r = 2$ on \gls{ris}-\gls{cc}. Nevertheless, \change{employing an OB-\gls{ris}-\gls{cc} can reduce the overhead by transmitting the \gls{ris} control packets through orthogonal resources.} On the other hand, the Algorithmic phase overhead is \emph{\gls{cc} independent} and \emph{transmission paradigm dependent}, being designed to achieve the goal of the specific paradigm regardless of the \gls{cc}. In the following, the overhead is evaluated.

\subsubsection{Initialization phase}
This phase starts with the initialization control packet sent on the \gls{ue}-\gls{cc} (INI-U) informing the \gls{ue} that the \gls{oce} procedure has started. In the \gls{ibcc} case, this is followed by the transmission of the INI-R packet to the \gls{risc} to notify the beginning of the procedure. A consequent \gls{tti} for feedback is reserved to notify back to the \gls{bs} if the INI-R packet has been received. If an \gls{obcc} is employed, no \gls{tti} needs to be reserved because the INI-R and its feedback are scheduled simultaneously since the INI-U packet relies on different resources (see Assumption~\ref{assu:ris-cc}).
The phase duration is $\tau\ini = T$ or $\tau\ini=3T$ with OB- or IB-\gls{ris}-\gls{cc}, respectively.

\subsubsection{Setup phase}
 After the optimization has run, a setup (SET-U) packet spanning one \gls{tti} is sent to the \gls{ue} notifying it to prepare to send the data; then, with an \gls{ibcc}, a \gls{tti} is used to send the SET-R packet containing the information of which configuration to load during the Payload phase; a further \gls{tti} is reserved for feedback. Again, if an \gls{obcc} is present, the SET-R and its feedback are scheduled at the same time as the SET-U packet but on different resources; therefore, no \glspl{tti} needs to be reserved for the SET-R and its feedback. Remark that the $\tau_s$ guard period must be considered by the \gls{ue} when transmitting the data to avoid being disrupted during the load of the configuration employed in the Payload. For simplicity of evaluation, we account for this guard period in the Setup phase duration, resulting in $\tau\set = \tau\ini + \tau_s$.

\subsubsection{Algorithmic phase}
This phase comprises the process of sending pilot sequences and the consequent evaluation of the configuration for the transmission. Regardless of the paradigm, we assume each pilot sequence spans an entire \gls{tti}, but the configuration's switching time must be considered a guard period. Therefore, the actual time occupied by a pilot sequence is $\tau_p \le T - \tau_s$ and the number of samples $p$ of every pilot sequence is given by p = $\left\lfloor \frac{T - \tau_s}{T_n} \right\rfloor$,
where $T_n$ is the symbol period in seconds. Assuming that the \gls{tti} duration and the symbol period are fixed, the \gls{ue} can compute the pilot length if it is informed about the guard period. The overall duration of the Algorithmic phase depends on the paradigm employed.

\paragraph{\gls{oce}} In this case, the Algorithmic phase starts with $C\oce$ \glspl{tti}; at the beginning of each of them, the \gls{risc} loads a different configuration, while the \gls{ue} transmits replicas of the pilot sequence. After all the sequences are received, the \gls{ce} process at the \gls{bs} starts, followed by the configuration optimization. The time needed to perform the \gls{ce} and optimization processes depends on the algorithm and the available hardware. To consider a generic case, we denote this time as $\tau_A = A T$.

\paragraph{\gls{bsw} fixed frame structure}
Similarly to the previous case, the Algorithmic phase starts with $C\bsw$ \glspl{tti}, at the beginning of which the \gls{risc} loads a different configuration, and the \gls{ue} transmits replicas of the pilot sequence. After receiving all sequences, the \gls{bs} selects the configuration as described in Section~\ref{sec:communication-paradigms:bsw}. The time needed to select the configuration is considered negligible. Thus, the Setup phase may start in the \gls{tti} after the last pilot sequence is sent.

\paragraph{\gls{bsw} flexible frame structure}
In this case, the number of \glspl{tti} used for the beam sweeping process is not known \emph{a priori} and depends on the measured \gls{snr}. However, to allow the system to react if the desired threshold is reached, a \gls{tti} is reserved for transmitting an acknowledgment (ACK-U) packet after each \gls{tti} used for pilot transmission. Hence, the number of \glspl{tti} needed is $2 c^\star - 1$, where $0< c^\star \le C\bsw$ is a random variable.

Accordingly, the Algorithmic phase duration is
\begin{equation} \label{eq:algorithmic-time}
    \tau\alg = 
    \begin{cases}
        (C\oce + A) T, \quad &\text{\gls{oce}}, \\
        C\bsw T, \quad &\text{\gls{bsw} fixed frame}, \\
        (2 c^\star - 1) T, \quad &\text{\gls{bsw} flexible frame}.
    \end{cases}
\end{equation}

\subsection{Reliability evaluation}
\label{sec:reliability}

\change{The reliability of the control packets depends on their informative content, the time reserved for their transmission, and the bandwidth of the \gls{cc}. With equal transmission time and bandwidth, transmitting a high informative packet is less reliable than a low informative packet; similarly, increasing the time reserved leads to higher reliability.
We account for this behavior via the outage probability of the $i$-the control packet intended to entity $k\in\{u,r\}$, which is given by}
\begin{equation} \label{eq:outagepe}
    p_i^{(k)} = \Pr\left\{ \log \left(1 + \Gamma_k \right) \le \frac{b_i^{(k)}}{\tau_{i}^{(k)} B_{k}} \right\}, \quad i =\{1,2\},
\end{equation}
where $i =1,2$ refers to the INI or SET packet, respectively; \change{$b_i^{(k)}$ is the amount of informative bits,} $\tau_{i}^{(k)}$ is the reserved time for transmission, and $B_k$ is the \gls{cc} bandwidth. Following the channel model in Section~\ref{sec:model},~\eqref{eq:outagepe} can be rewritten as
\begin{equation} \label{eq:outagepe2}
    p_i^{(k)} = 1 - \exp\left[- \frac{1}{\lambda_k} \left(2^{b_i^{(k)} / \tau_{i}^{(k)} / B_{k}} - 1 \right) \right].
\end{equation}
Plugging~\eqref{eq:outagepe2} into~\eqref{eq:pcc}, the correct control probability is
\begin{equation} \label{eq:pcc2}
\begin{aligned}
p_\mathrm{cc} =& \exp\left[ \frac{1}{\lambda_u} \left( 2 - \sum_{i=1}^2 2^{b_{i}^{(u)} / \tau_{i}^{(u)} / B_u}\right)\right] \times \\ &\exp\left[ \frac{1}{\lambda_r} \left( 2 - \sum_{i=1}^2 2^{b_{i}^{(r)} / \tau_{i}^{(r)} / B_r} \right) \right].    
\end{aligned}
\end{equation}
\change{Remark that the informative content $b_i^{(k)}$ and the reserved time $\tau_i^{(k)}$ depend on the control packet and the communication paradigm employed due to the need to communicate different control information. Hence, different paradigms require different values of the average \gls{snr} of the \gls{ue}-\gls{cc} $\lambda_u$ and the \gls{ris}-\gls{cc} $\lambda_r$ to obtain the same value of $p_\mathrm{cc}$. In practice, the minimum $\lambda_u$ and $\lambda_r$ required to obtain the target correct control probability give a measure of the complexity of the decoding process.}
In the following, we compute $\tau_i^{(k)}$ and $b_i^{(k)}$ for the cases under investigation.

\subsubsection{Reserved time for control packets}\label{sec:usefultime}
Each control packet spans an entire \gls{tti} following the data frame. However, the actual transmission time $\tau_{i}^{(k)}$, \emph{i.e.}, the time in which informative bits can be sent without risk of being disrupted, depends on the \gls{ris} switching time. As discussed in Section~\ref{sec:overhead}, a guard period $\tau_s$ must be considered if the \gls{risc} loads a new configuration in that \gls{tti}. Following the frame structure of Fig.~\ref{fig:data-frames}, INI-R and SET-R packets can use the whole \gls{tti}, while INI-U packets need the guard period. The SET-U control packet does not employ the guard period under the \gls{oce}, as long as $A \ge 1$. The guard period is needed for the \gls{bsw} paradigm. Hence, the transmission time of the control packets intended for the \gls{ue} is $\tau_1^{(u)} = T - \tau_s$ for all paradigms, and $\tau_2^{(u)} = T - \tau_s$ for \gls{bsw} and $\tau_2^{(u)}$ for \gls{oce},
while the time reserved for the control packets intended for the \gls{risc} is $\tau_{1}^{(r)} = \tau_{2}^{(r)} = T$.

\subsubsection{Control packet content}
\label{sec:bits}
\begin{figure}
    \centering
    \includegraphics[height=1.3cm]{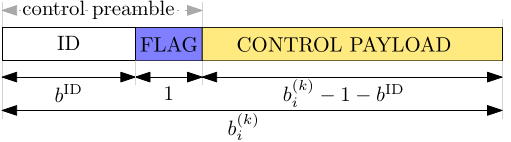}
    \caption{General control packet structure.}
    \label{fig:packet-structure}
    \vspace{-0.5cm}
\end{figure}

Without loss of generality, we can assume a common structure for all the control packets, comprising a control preamble and a control payload as depicted in Fig.~\ref{fig:packet-structure}. The preamble comprises $b^\mathrm{ID}$ bits representing the \emph{unique identifier (ID)} of the destination entity in the network and a single bit flag specifying if the packet is a INI or a SET one. From the preamble, the entity can understand if the control packet is meant to be decoded and how to interpret the control payload, whose informative bits depend on the kind of control packet and on the communication paradigm considered.

\paragraph{\gls{oce}}
To initialize the overall procedure at the \gls{ue}, the payload of the INI-U packet must contain the length of the frame $\tau$, the cardinality of the set $C\oce$, and the guard time $\tau_s$. To simplify the data transmission, the frame duration can be notified through an (unsigned) integer $b^\mathrm{frame}$ containing the number of total \glspl{tti} $\lceil \tau / T \rceil$ set for the frame. Similarly, we can translate the guard time into an unsigned integer representing the number of guard symbols $\lceil \tau_s / T_n \rceil$ to send $b^\mathrm{guard}$ bits. Finally, another integer of $b^\mathrm{conf}$ bits can be used to represent the cardinality $C\oce$ and to notify it to the \gls{ue}. Its minimum value is $b^\mathrm{conf} = \lfloor \log_2(C) \rfloor$, where $C$ is the total number of configurations stored in the common codebook.
Similarly, the payload of the INI-R packets needs to contain the information of the length of the frame $\tau$, and the \emph{set} of configuration  $\mc{C}\oce$ to switch through.
The former uses the same $b^\mathrm{frame}$ bits of the INI-U packet. To encode the latter, $b^\mathrm{conf}$ bits are used to identify a single configuration in the common codebook, and thus, $C\oce b^\mathrm{conf}$ needs to be transmitted to the \gls{risc}, one per desired configuration.
Regarding the Setup phase, the payload of the SET-U contains only the chosen \gls{se} of the communication $\eta\oce$. This can be encoded similarly to the \gls{mcs} in the 5G standard~\cite{3gpp:rel15}: a table of predefined values indexed by $b^\mathrm{SE}$ bits. The payload of the SET-R must contain the optimal configuration $\bm{\phi}_\star$, that is, a phase-shift value for each element. Without loss of generality, we denote by $b^\mathrm{quant}$ the number of bits used to control each element, \emph{i.e.}, the level of quantization of the \gls{ris}~\cite{EURASIP_RIS}. Hence, the overall number of informative bits is 
$N b^\mathrm{quant}$.
To summarize:
\begin{equation} \label{eq:bits:oce}
    b_{i}^{(k)}\hspace{-0.2mm} = \hspace{-0.2mm} b^\mathrm{ID} \hspace{-1mm}+\hspace{-0.5mm} 1 \hspace{-0.5mm}+\hspace{-1mm}
    \begin{cases}
        b^\mathrm{frame} + b^\mathrm{guard} + b^\mathrm{conf}, \quad &k=u, \, i= 1, \\
        b^\mathrm{frame} + C\oce b^\mathrm{conf}, \quad &k= r, \, i=1, \\ 
        b^\mathrm{SE}, \quad &k= u, \, i=2,\\
        N b^\mathrm{quant}, \quad &k= r, \, i=2.
    \end{cases}
\end{equation}

\paragraph{\gls{bsw}}
The payload of the Initialization packets follows the same scheme used for the \gls{oce} paradigm. The INI-U packet contains the length of the frame $\tau$, the cardinality of the set $C\bsw$, and the guard time $\tau_s$ in the (unsigned) integers $b^\mathrm{frame}$, $b^\mathrm{guard}$, and $b^\mathrm{conf}$, respectively.
The payload of the INI-R packets contains the information of the length of the frame $\tau$, and the \emph{set} of configuration  $\mc{C}\bsw$ to switch through, encoded in the (unsigned) integers $b^\mathrm{frame}$ and $C\bsw b^\mathrm{conf}$, respectively.
Instead, the Setup contains different information. In particular, the payload of the SET-U is empty, according to the fixed rate transmission used by this paradigm. The payload of the SET-R contains the configuration $c^\star$ encoded by the same $b^\mathrm{conf}$ bits, representing an index in the common codebook. 
To summarize, the packet length is:
\begin{equation} \label{eq:bits:bsw}
    b_{i}^{(k)}\hspace{-0.2mm} = \hspace{-0.2mm} b^\mathrm{ID} \hspace{-1mm}+\hspace{-0.5mm} 1 \hspace{-0.5mm}+\hspace{-1mm}
    \begin{cases}
        b^\mathrm{frame} + b^\mathrm{guard} + b^\mathrm{conf}, \quad &k=u, \, i= 1,\\
        b^\mathrm{frame} + C\bsw b^\mathrm{conf}, \quad &k= r, \, i=1, \\ 
        0, \quad &k= u, \, i=2, \\
        b^\mathrm{conf}, \quad &k= r, \, i=2.
    \end{cases}
\end{equation}
\change{Remark that the informative content of the \gls{bsw} packets is lower or equal to the one of \gls{oce}, leading the former to be more reliable than the latter.}

\section{Numerical Results and Discussion} \label{sec:results}

\begin{table}[bt]
    \centering
    \caption{Simulation parameters; $\nu$ represents the speed of light.}
    \footnotesize
    \begin{tabular}{lcc}
    \toprule    
    \multicolumn{3}{c}{\textbf{Scenario}}  \\ 
    \midrule
    Scenario side & $D$ & $20$~m  \\
    BS position & $\mb{x}_b$ & $(25, 5, 5)\T$~m \\
    RIS element spacing & $d$ & $\nu / f_d / 2$ \\
    Number of RIS elements & $N$ & 100  \\
    \midrule     
    \multicolumn{3}{c}{\textbf{Communication}} \\
    \midrule
    \gls{dc} frequency & $f_d$ & 3 GHz\\ 
    \gls{dc} bandwidth & $B_d$ & $180$~kHz\\
    \gls{ue} transmit power & $\rho_u$, $\rho_b$ & 24 dBm \\
    \gls{ue}/\gls{ris} noise power & $\sigma_u^2, \sigma_r^2$ & $-94$~dBm\\
    \midrule
    \multicolumn{3}{c}{\textbf{Paradigms}} \\ 
    \midrule     
     Codebooks cardinality & $C\oce$, $C\bsw^\mathrm{fix}$, $C\bsw^\mathrm{fle}$  & $N$, $\lceil N/3 \rceil$, $N$ \\
     Overall frame duration & $\tau$ & [10, 200] ms \\
     \gls{tti} duration & $T$ & 0.5 ms \\
     Guard period & $\tau_s$ & 50 $\mu$s \\
     Pilot sequence length & $p$ & 1 \\
     \gls{ce} duration in \glspl{tti} & $A$ & 5 \\
     \midrule     
     \multicolumn{3}{c}{\textbf{Control packet content}} \\
     \midrule
     ID \& Configuration bits & $b^\mathrm{ID}$, $b^\mathrm{conf}$ & 8\\
     \glspl{tti} \& Guard period bits & $b^\mathrm{frame}$, $b^\mathrm{guard}$ & 16\\     
      \gls{se} table bits & $b^\mathrm{SE}$ & 6\\
      \gls{ris} quantization level bits& $b^\mathrm{quant}$ & 2\\     
      \bottomrule
    \end{tabular}
    \label{tab:params}
    \vspace{-0.3cm}
\end{table}

This section presents the performance evaluation of the communication paradigms under consideration.\footnote{The simulation code for the paper is available at \url{https://github.com/lostinafro/ris-control}.} The parameters set for the simulations are given in Table~\ref{tab:params}, if not otherwise specified. The scenario is tested through Monte Carlo simulations. Concerning the scenario described in Section~\ref{sec:model}, we consider that the \gls{bs} and \gls{ris} positions $\mb{x}_b$ and $\mb{x}_r=(0,0,0)\T$ are kept fixed, while the \gls{ue} position, $\mb{x}_u$, changes at every simulation according to a uniform distribution having limits $(-D/2, 0, 0)\T$ and $(D/2, D, -D)\T$. When referring to average performance, we implicitly assume averaging over different \gls{ue} positions and noise realizations.
The channel coefficients of the \gls{dc} are evaluated considering the \gls{los} component of $\mb{h}_d$ and $\mb{g}_d$ following the model of~\cite[Section II]{albanese2022marisa}. Note that the \gls{ris}-\gls{obcc} uses a different operating frequency and bandwidth w.r.t. the \gls{dc} ones, while, for the \gls{ibcc}, we have $f_r = f_d$ and $B_r =  5 B_d$. The \gls{ue}-\gls{cc} has operating frequency $f_u = f_d$ and bandwidth $B_u = 5 B_d$.
The overall frame duration $\tau$ reflects the coherence time of the channel: a low $\tau$ represents a high mobility environment with low coherence time, and vice versa. The \gls{tti} duration is set according to the half of the subframe duration in the \gls{ofdm} 5G NR standard~\cite{3gpp:rel15}. For the \gls{oce} paradigm, the channel estimation codebook $\mc{C}\oce$ is designed from the \gls{dft}, as described in Section~\ref{sec:communication-paradigms:oce}. For the sake of simplicity, the same configurations are used in the beam sweeping codebook $\mc{C}\bsw$. In particular, the codebook used by the \gls{bsw} with flexible frame structure is $\mc{C}\bsw^\mathrm{fle} = \mc{C}\oce$, while the one used by the \gls{bsw} with fixed frame structure uses one every three configurations, to take advantage of the possible lower overhead. Next, we divide the results into two parts: evaluating the paradigms performance under error-free \glspl{cc} and investigating the impact of \glspl{cc} reliability.

\subsection{Paradigms performance (error-free \glspl{cc})}
\begin{figure}[!t]
    \centering
    \input{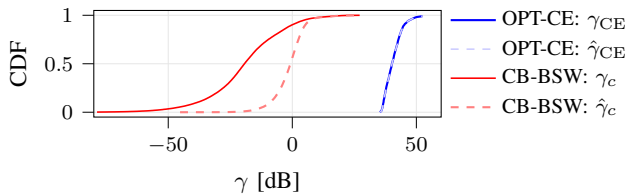}
    \vspace{-0.2cm}
    \caption{\gls{cdf} of the actual and estimated \gls{snr} for the communication paradigms under investigation.}
    \label{fig:snrcdf}  
    \vspace{-0.5cm}
\end{figure}

Fig.~\ref{fig:snrcdf} depicts the \gls{cdf} of the actual and estimated \glspl{snr} to give some insight into the impact of the possible algorithm errors. From the figure, it can be inferred that the noise impact on the \gls{snr} estimation is generally negligible for the \gls{oce} paradigm. This finding is justified by the fact that the power of the noise influencing the measurement is proportional to $1/C\oce$, where $C\oce \ge N$ (see Section~\ref{sec:communication-paradigms:oce}). On the other hand, when employing the \gls{bsw} paradigm, the \gls{snr} is measured per each configuration, and the resulting noise has a higher impact on the estimation. Moreover, it can be observed that the \gls{snr} of the \gls{bsw} extends to very low values, while the minimum target \gls{kpi} needs to be set to provide a non-negligible and supportable \gls{se} (at least higher than $-13$ dB to reach the minimum \gls{se} value $0.0586$ of 5G NR~\cite[Table 5.1.3.1-3]{3gpp:nr}). Therefore, whatever reasonable target \gls{kpi} is chosen, it will produce a relatively high outage probability. In summary, the \gls{oce} paradigm is inherently more robust to algorithmic errors than the \gls{bsw} one.

\begin{figure}[tb]
    \centering
    \input{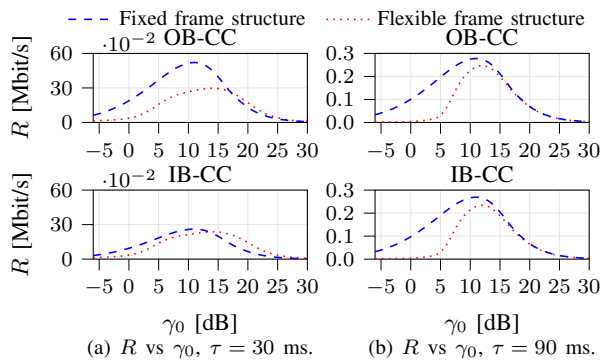}
    \vspace{-0.6cm}
    \caption{Analysis of the target \gls{snr} for the \gls{bsw} paradigm (note the different scale in Fig.~\ref{fig:kpi:a}).}
    \label{fig:kpi}
    \vspace{-0.3cm}
\end{figure}

To provide a fair comparison between the paradigms, we evaluate the optimal target \gls{snr} $\gamma_0$ used as relevant \gls{kpi} for the \gls{bsw} paradigm. 
Fig.~\ref{fig:kpi} shows the average goodput $R$ achieved as a function of the target \gls{snr}, under different kinds of \gls{cc} and for different $\tau$ values. We note that the optimal $\gamma_0$ depends on the frame structure chosen, while it does not depend on the kind of \gls{cc}. 
Moreover, the duration $\tau$ influences negligibly the optimal $\gamma_0$ in the flexible structure, being approximately $13.8$ dB for $\tau=30$ ms and $12.4$ dB for $\tau=90$ ms for the flexible frame structure, while approximately $10.9$ dB for any value of $\tau$ for the fixed frame structure.
We remark that the selection of the target \gls{kpi} is also scenario-dependent; hence, this procedure should be performed during the deployment of the \gls{ris}.

Using the optimal target \gls{snr}, we now compare the performance of the two paradigms. 
Fig.~\ref{fig:goodput:a} illustrates $R$ as a function of the overall frame duration. Again, the impact of the kind of \gls{cc} on $R$ is negligible. The main advantage of the \gls{bsw} approach is the possibility of providing a non-null transmission rate even in the presence of a lower coherence block ($< 60$ ms), while the \gls{oce} needs a longer time horizon to obtain the \gls{csi} and perform the Payload phase ($\ge 60$ ms). On the other hand, as long as the time horizon is sufficiently long ($\tau \ge 75$ ms), the \gls{oce} paradigm outperforms the \gls{bsw} one.
In Fig.~\ref{fig:goodput:b}, we show the \gls{cdf} of the goodput for $\tau = 60$ ms. For this frame duration, the kind of \gls{cc} influences the performance of the \gls{oce} paradigm, while its impact is less predominant on the \gls{bsw} performance. As expected, the \gls{ibcc} performs worse due to its increased overhead. Nevertheless, for the \gls{bsw}, approximately half of the transmissions have null goodput because of algorithmic errors, while the \gls{oce} provide a non-null goodput for all values, corroborating the results of Fig.~\ref{fig:snrcdf}.

\begin{figure}[bt]
    \centering
    \input{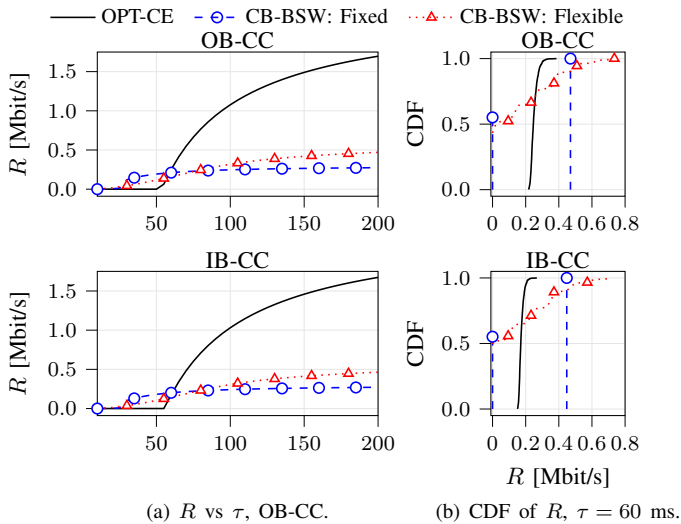}    
    \vspace{-0.6cm}
    \caption{Analysis of the goodput performance.}
    \label{fig:goodput}
     \vspace{-0.3cm}
\end{figure}

\subsection{Impact of the \glspl{cc} reliability}
\begin{figure}[tb]
    \centering
    \input{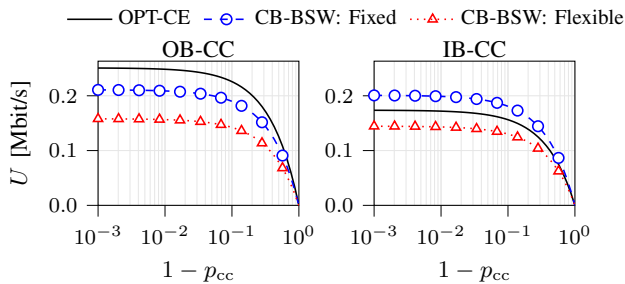}
    \caption{Analysis of the utility function vs. erroneous control probability for \change{$\tau = 60$ ms}.}
    \label{fig:utility}
    \vspace{-0.5cm}
\end{figure}

Fig.~\ref{fig:utility} demonstrates the average utility~\eqref{eq:utility} as a function of the erroneous control probability $1 - p_\mathrm{cc}$, for $\tau = 60$ ms and both kinds of \gls{cc}. The results of the transmission paradigms are in line with the ones presented in Fig.~\ref{fig:goodput}. The \gls{cc} reliability influences significantly the performance when $1 - p_\mathrm{cc} \le 0.05$, \emph{i.e.}, $p_\mathrm{cc} \le 0.95$. 
To consider a conservative case, we set a target reliability to be $\bar{p}_\mathrm{cc} = 0.99$, and we study the minimum average \glspl{snr} $\lambda_u$ and $\lambda_r$ providing such reliability, following the control packet content given in Table~\ref{tab:params}. Fig.~\ref{fig:reliability:a} shows the achieved $p_\mathrm{cc}$ for the \gls{obcc} as a function of $\lambda_u$ only, according to the assumption of error-free \gls{ris}-\gls{cc} in the \gls{obcc} case. With this kind of \gls{cc}, the probabilities of correct control achieved by \gls{oce} and \gls{bsw} have negligible differences, and $\bar{\lambda}_u = 10.5$ dB is enough to provide the target correct control probability in both cases. Fig.~\ref{fig:reliability:b} shows the $p_\mathrm{cc}$ as a heatmap function of $\lambda_r$ and $\lambda_u$ for the \gls{ibcc} case. Only the region satisfying the $p_\mathrm{cc} \ge 0.99$ is colored, and the minimum \glspl{snr} $\bar{\lambda}_r$ and $\bar{\lambda}_u$ needed are sketched. 
\change{The \gls{oce} needs higher \glspl{snr} than \gls{bsw} due to the higher information content of the control packets of the former. In particular, the need to transmit the phase-shift of each \gls{ris} element in the \gls{oce} paradigm strongly impacts the overall reliability (see~\eqref{eq:bits:oce}). We remark that the performance provided by the \glspl{cc} should be satisfied \emph{simultaneously} to achieve the desired control reliability.}
\vspace{-0.3cm}

\begin{figure}[bht]
    \centering
    \input{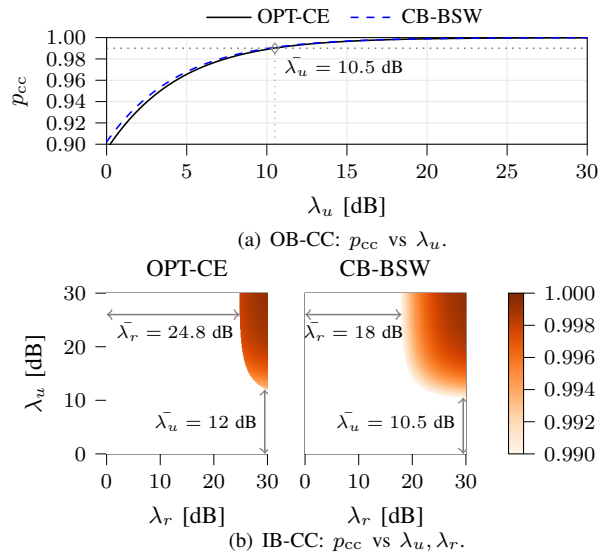} 
    \vspace{-0.2cm}
    \caption{Impact of the \gls{cc} \gls{snr} on the reliability performance.}
    \label{fig:reliability}
    \vspace{-0.5cm}
\end{figure}

\section{Toward more elaborate control designs} 
\label{sec:extension}
\change{The presented performance analysis framework paves the way for the control signaling design and quantification of more sophisticated \gls{ris}-empowered wireless systems. It can be applied, for example, to multi-user wideband/\gls{ofdm} communications~\cite{Huang2019,risTUTORIAL2020}, by accounting for the subcarrier allocation of the different control and payload messages. For this system setup, the Algorithmic phase needs to also consider the resource allocation process, whose output should be signaled to the \glspl{ue} through a specific design of the Setup phase. In addition, the current framework, by omitting, merging, or repeating some of its general phases, can set the basis for the control design in \gls{ris}-assisted networks with a multi-antenna \gls{bs} and multi-antenna \glspl{ue}, and smart wireless environments with multiple, possibly machine learning orchestrated, \glspl{ris}~\cite{RIS_pervasive_ML}, as well as shared \glspl{ris} among multiple communications pairs~\cite{EURASIP_RIS}. Of late interest are also multi-functional \glspl{ris}~\cite{Tsinghua_RIS_Tutorial}, and especially those possessing sensing capabilities~\cite{HRIS_VTM}, which may provide higher flexibility for efficient control signaling designs~\cite{croisfelt2023orchestration}.}

\change{We next elaborate in more detail on the case where the \gls{bs} is equipped with multiple antennas and there exists the possibility of a weak direct link between itself and the \gls{ue}. For the \gls{oce} communication paradigm, the \gls{ris} configuration and the \gls{bs} combiner can be jointly optimized~\cite{risTUTORIAL2020}, at the cost, of course, of a larger \gls{ce} overhead and complexity, as well as larger algorithmic complexity. It is noted, however, that the increased beamforming gain from the multiple \gls{bs} antennas might lead to cases where the \gls{bs}-\gls{ue} link is satisfactory, implying that the \gls{ris} deployment can be avoided, reducing the control overhead. For this mode to be realized, the operation protocol needs to enable, for example, the separate estimation of the \gls{ue}-\gls{ris}-\gls{bs} and \gls{ue}-\gls{bs} channels, via activation/deactivation of the \gls{ris} panel, as well as a relevant action during the Initialization phase.}   
\change{There exist various modes of operation when the \gls{bsw} paradigm is adopted. One is to perform \gls{BSW} at the \gls{bs} during the Initialization phase, together with \gls{BSW} at the \gls{ris}, again at a cost of a larger overhead for both the fixed and flexible frame structures. Alternatively, to reduce the control signaling overhead, the \gls{bs} combiner can be designed to solely match its channel with the \gls{ris}, or that with the \gls{ue} if the \gls{ris} can be avoided, as discussed in the \gls{oce} paradigm. One way to achieve the former is to capitalize on the common assumption that the \gls{ris} is placed such that there exists a strong line-of-sight with the BS~\cite{RIS_security_mal}. In this way, the \gls{bs} may adopt the reception configuration closest to maximal ratio combining. When the \gls{ris} can be avoided, the \gls{bs} combiner can be similarly designed, now for the channel towards the \gls{ue} - this operational mode can be decided similarly to the respective \gls{oce} case.}

\change{It is finally noted that the two presented communication paradigms, namely \gls{oce} and \gls{bsw}, can be combined to devise additional signaling schemes. One example is presented in~\cite{Jamali2022}, where the \gls{bsw} is performed to set the \gls{ris} configuration when the link budget falls below a certain threshold, and then \gls{oce} follows to set the BS combiner and/or the \gls{ue} beamformer, treating the configured \gls{ris} as an unknown scatterer.}

\section{Conclusions} \label{sec:conclusions}
In this paper, we proposed a general framework of four phases (Initialization, Algorithmic, Setup, and Payload) to evaluate \gls{ris}-enabled communication performance while addressing the impact of control and signaling procedures. The data exchange and the frame structure for two different communication paradigms, namely \gls{oce} and \gls{bsw}, were detailed. The performance of the paradigms was analyzed employing a utility function that considers the overhead generated by its various phases, the possible errors coming from the Algorithmic phase, and the impact of losing control packets needed for signaling purposes. Moreover, we particularized the performance evaluation for two kinds of \glspl{cc} connecting the decision maker and the \gls{risc} --  \gls{ibcc} and \gls{obcc} --, showcasing the minimum performance needed to obtain the desired control reliability. \change{Possible extensions of the proposed framework 
for more sophisticated scenarios of interest were discussed. Together with those extensions, in the future, we intend to study the impact of synchronization errors on the frame level and in the PHY-layer resources.}




\bibliographystyle{IEEEtranNoURL}
\bibliography{IEEEabbr,bib}

\end{document}